\documentclass[11pt]{article}

\usepackage[margin=1in]{geometry}
\usepackage[hyphens]{url}
\usepackage{graphicx}
\usepackage{booktabs}
\usepackage{amsmath,amssymb,amsfonts}
\usepackage{amsthm}
\usepackage{algorithm}
\usepackage{algorithmic}
\usepackage{multirow}
\usepackage{array}
\usepackage{microtype}
\usepackage{xcolor}
\usepackage[numbers,sort&compress]{natbib}
\usepackage{hyperref}
\usepackage[nameinlink,noabbrev]{cleveref}

\hypersetup{
  colorlinks=true,
  linkcolor=blue!55!black,
  citecolor=green!35!black,
  urlcolor=blue!60!black,
  pdftitle={Stop When Certain: 74x Cheaper Agent Evaluation with Certified Anytime-Valid Stopping in Imperfect-Information Games},
  pdfauthor={Boning Li and Yu Chen and Longbo Huang}
}
\newtheorem{proposition}{Proposition}
\newtheorem{theorem}{Theorem}
\newtheorem{lemma}{Lemma}
\newtheorem{corollary}{Corollary}
\newtheorem{assumption}{Assumption}
\theoremstyle{remark}
\newtheorem{remark}{Remark}

\newcommand{\F}{\mathcal{F}}
\newcommand{\E}{\mathbb{E}}
\newcommand{\Prob}{\mathbb{P}}
\newcommand{\Var}{\mathrm{Var}}
\newcommand{\hw}{w}
\newcommand{\stope}{\tau}
\newcommand{\loga}{\log\tfrac{2}{\alpha}}
\newcommand{\clip}{\mathrm{clip}}
\newcommand{\vstar}{v^{\star}}
\newcommand{\sstar}{\sigma_{\star}}

\graphicspath{{figures/}}

\title{\mbox{AV-AIVAT}: $74\times$ Cheaper Agent Evaluation\\
with Certified Anytime-Valid Stopping in Imperfect-Information Games}

\author{Boning Li\thanks{IIIS, Tsinghua University. Email: \texttt{li-bn22@mails.tsinghua.edu.cn}.}
\and Yu Chen\thanks{IIIS, Tsinghua University.}
\and Longbo Huang\thanks{IIIS, Tsinghua University. Email: \texttt{longbohuang@tsinghua.edu.cn}. Corresponding author.}}

\date{August 7, 2026}

\begin{document}
\maketitle

\begin{abstract}
Deciding which of two agents is stronger means playing games until skill
outweighs luck, and every game costs money, model inference, or expert time.
Since the number of games needed is unknown, fixed-budget evaluations either
keep paying after the result is settled or stop before the agents can be told
apart, while naive optional stopping with an ordinary confidence interval
invalidates the stated level.  We make such an evaluation stop as soon as its
evidence suffices, with the guarantee intact.  The Action-Informed Value
Assessment Tool (AIVAT) reduces variance in imperfect-information games through
conditional mean-zero corrections, by a median $54\times$ across 15
LLM agent configurations spanning 71{,}439 paired Heads-Up
No-Limit Hold'em (HUNL) hands, but does not say when to stop.  We combine AIVAT
with continuously monitored Confidence Sequences (CSs) into anytime-valid
\mbox{AIVAT} (AV-AIVAT), whose online value model learns only from past games so
that no game scores its own correction.  At the nominal 95\% level and a target
precision of $\pm1$ Big Blind (BB), raw outcomes need a median $74\times$ as
many hands as AIVAT-corrected outcomes to stop under the Asymptotic CS (AsympCS).  Exact finite-sample certification uses the
Empirical-Bernstein CS (EB-CS), which needs
an independently justified bound on corrected payoffs.  We establish such a
bound structurally for Leduc hold'em and characterize a width floor set by the
CS's bet cap and that bound, which governs how much of a variance gain becomes
earlier stopping; the descriptive HUNL EB-CS runs show a median $1.37\times$
stopping-time ratio.  AV-AIVAT turns variance reduction into efficient,
auditable early stopping while separating asymptotic screening from exact
certification, so an evaluation can stop the moment its evidence suffices and
hand a third party everything needed to recheck the verdict at that very
stopping time.

\end{abstract}

\section{Introduction}
\label{sec:intro}

Interactive agent evaluation is often both costly and noisy.  PokerSkill, a
framework for deploying LLM agents in complete poker
games, spends \$0.07--\$0.30 of inference per hand
\citep{li2026pokerskill}.  Human matches are no cheaper: Libratus played
120{,}000 hands against professionals over 20 days for a \$200{,}000 prize pool,
winning by $147$~mbb/hand at $99.98\%$ significance \citep{brown2018superhuman}.
The hand count was fixed when the match was designed, so the full schedule was
played and paid for however early the margin became decisive.  Because the
number of games needed is rarely known in
advance, the usual response is a fixed budget, which either overruns the evidence
or ends before a useful conclusion is possible.  Watching an ordinary
fixed-sample interval instead and stopping at the first favorable look
invalidates nominal coverage and inflates false positives.  The problem is
therefore to stop when the evidence is sufficient while preserving a stated
statistical guarantee.

Imperfect-information games (IIGs) provide a demanding setting for sequential
agent evaluation: outcomes have high variance, interactions may be costly, and
information available to an evaluator changes within each game.  Poker milestones
from solved limit hold'em to superhuman multiplayer play were all declared from such a stream
\citep{bowling2015heads,moravvcik2017deepstack,brown2018superhuman,brown2019superhuman}.
The action-informed value assessment tool (AIVAT) conditionally reduces outcome
variance using zero-mean corrections derived from a value function
\citep{burch2018aivat}.  Confidence sequences (CSs) solve a complementary
problem: they retain their stated guarantee under continuous monitoring and
optional stopping \citep{howard2021time,grunwald2024proposer}.
Classical AIVAT freezes its value function, while a CS on raw outcomes can still
require many costly interactions.  We combine them into anytime-valid AIVAT
(AV-AIVAT), a general statistical interface for evaluating agents in IIGs.

AV-AIVAT raises three practical questions.  Can the value model improve during
evaluation without letting the current game influence its own correction?  When
does lower variance actually reduce the number of games needed?  What must be
released so that others can verify an early-stopping claim?

\textbf{First, online learning is valid under a pre-action conditional-kernel
interface.}  At each correction node, the evaluator must know the action's
conditional kernel given the information available immediately before that
action, and any decision to enable the correction must already be fixed at that
time.  The value function $v_t$ may use games $1{:}t{-}1$, but not game $t$.
These conditions keep the outgoing action and the later trajectory from
selecting their own correction, preserving the conditional mean-zero law
while the value model sharpens as the evaluation proceeds.  Under additional
variance-adaptive width and sublinear variance-regret assumptions, the online
learner has no asymptotic delay relative to a deterministic oracle-width
benchmark (Theorem~\ref{thm:arx-variance-regret}), and across two mismatched
controlled profiles it recovers 77--79\% of the frozen-to-oracle variance gap.

The two confidence sequences build on this interface with different additional
requirements, so the paper states their guarantees separately.  The EB-CS attains
exact finite-sample validity once a data-independent almost-sure bound on the
corrected payoff is supplied (Proposition~\ref{prop:arx-predictable-validity}),
and we derive and exhaustively check such a bound in Leduc.  The asymptotic CS
(AsympCS) instead needs limiting variance conditions
(Proposition~\ref{prop:arx-asympcs-validity}); a locked replication-level split
measures its observed finite-horizon exclusion rate.

\textbf{Second, a width floor governs how much variance reduction becomes
earlier stopping.}  Our heads-up no-limit hold'em (HUNL) corpus contains
71{,}439 paired hands from 15
PokerSkill/LLM-agent configurations in fixed-opponent multi-hand runs
\citep{li2026pokerskill,provost2026gto}, with both per-hand raw and
AIVAT-corrected payoffs supplied by the evaluation platform.  AIVAT reduces
variance by a median $54\times$.  At $\alpha=0.05$, for a half-width of one big
blind (BB), paired replay yields a median raw-to-AIVAT stopping-time ratio of 74
under AsympCS, compared with 1.37 under the descriptive EB-CS execution.  The
difference has a clean explanation: in the bet-capped EB-CS, the declared payoff
bound imposes a deterministic width floor $4B\log(2/\alpha)/t$
(Lemma~\ref{lem:arx-bet-floor}), so once that floor binds, precision is set by
the bound rather than by observed variance, and the resulting three-regime
expression is an order-level design guide.

\textbf{Third, a fast evaluation claim can still be rechecked by others.}
Rechecking requires the corrected outcomes through the reported stopping time,
together with the stopping rule, target, confidence level, declared bound, and CS
settings; these released objects suffice to reconstruct the claim.  The same
time-uniform property makes continuous monitoring safe: in a rolling-leaderboard
simulation, monitoring an ordinary fixed-sample interval and stopping at the
first favorable look yields 61\% false positives under the null, whereas
recomputing the CS at that reported stopping time overturns essentially all such
claims.  When AsympCS continues under its own stopping rule, it later detects the
injected 0.1 and 0.2 BB effects in 81\% and 100\% of entries by 5,000 hands.

\paragraph{Contributions.}
Our contributions are:
\begin{itemize}
\item We introduce AV-AIVAT, a predictable AIVAT interface that exposes the
  conditional action kernel and fixes correction enablement before the action is
  observed.  Past-only value updates preserve the correction's conditional
  mean-zero property, while separate validity statements distinguish the exact
  bounded EB-CS certificate from the efficient asymptotic CS.
\item We establish how a bet-capped EB-CS converts variance reduction into
  earlier stopping.  A deterministic width floor induced by the declared payoff
  bound yields a three-regime, order-level design guide, while the
  variance-regret result shows when a past-only online value learner approaches
  the stopping efficiency of an oracle value function.
\item We propose a release protocol that makes an early-stopping claim directly
  reconstructible.  Publishing the corrected prefix together with the stopping
  rule and index, first-look and censoring metadata, declared bound, and locked
  CS settings lets a third party recompute the reported intervals at the exact
  stopping time without requiring the raw payoff stream.
\end{itemize}

\section{Related Work}
\label{sec:related}

Our work lies at the intersection of costly interactive evaluation, variance
reduction, and anytime-valid inference. The relevant methods solve different
parts of the problem: benchmark protocols define the stream, control variates
make it less noisy, and sequential inference determines what can be concluded
under continuous monitoring. Our contribution concerns the interface among
these components rather than a replacement for any one of them.

\paragraph{Costly interactive evaluation.}
Unlike static benchmark scoring, interactive evaluation consumes resources with
every additional game. This is explicit for LLM agents: PokerSkill reports
approximately \$0.07--\$0.30 of model inference per heads-up no-limit hold'em
(HUNL) hand, making statistical
sample efficiency a direct determinant of cost and latency
\citep{li2026pokerskill}. Human comparisons likewise require sustained expert
participation. DeepStack played 44{,}852 games against 33 professionals
\citep{moravvcik2017deepstack}; Libratus used a 20-day,
120{,}000-hand match with a \$200{,}000 prize pool
\citep{brown2018superhuman}; and Pluribus used 10{,}000-hand protocols with
participation and performance incentives \citep{brown2019superhuman}. These
fixed-scale evaluations established landmark results, but their budgets do not
supply a statistically valid rule for terminating a future evaluation as soon
as sufficient evidence accumulates.

\paragraph{LLM agents and poker benchmarks.}
PokerBench evaluates poker reasoning and decisions in curated situations
\citep{zhuang2025pokerbench}, while the GTO Wizard Benchmark evaluates agents
through gameplay against a strong common opponent \citep{provost2026gto}.
PokerSkill instantiates LLMs as agents in complete HUNL play without training or
solver calls at deployment \citep{li2026pokerskill}. These efforts address what
to evaluate and how to generate strategically meaningful interactions. We ask a
complementary question: given the costly payoff stream produced by such a
protocol, how can an evaluator reduce its noise, monitor it continuously, stop,
and release enough evidence for the stopping claim to be checked?

\paragraph{What the evaluated stream compares.}
The agents whose payoffs an evaluator observes are typically produced by
CFR and its descendants
\citep{zinkevich2007regret,lanctot2009monte,tammelin2014solving}, accelerated by
discounting \citep{brown2019solving}, warm starting \citep{brown2016strategy}, and
regret-based pruning \citep{brown2015regret,brown2017reduced}, with function
approximation replacing tabular regrets at scale
\citep{brown2019deep,brown2020combining,fu2021actor,mcaleerescher}. Their strength
depends on abstraction quality
\citep{waugh2009abstraction,johanson2013evaluating,li2024rl,li2026effective} and on
real-time refinement of the subgame actually reached
\citep{ganzfried2015endgame,moravcik2016refining,brown2017safe,brown2018depth,kovavrik2023value}.
Where a game is small enough, best-response computation inside the abstraction
gives an exact quality measure \citep{johanson2011accelerating}; at HUNL scale
\citep{johanson2013measuring} that option is unavailable, and competition-grade
agents \citep{jackson2013slumbot,brown2016baby} are ranked by played hands
instead. The same holds beyond poker, where variance-reduced self-play evaluation
is used in other large imperfect-information card games \citep{zha2021douzero}.
Our contribution concerns the statistics of that ranking stream, not the solvers
that generate it.

\paragraph{Variance reduction for game and agent evaluation.}
Poker evaluation has a long history of reducing outcome noise. Prior approaches
include duplicate designs and optimal low-variance unbiased estimators
\citep{zinkevich2006optimal}, DIVAT
\citep{billings2006tool,kan2007postgame}, learned value functions
\citep{white2009learning}, strategy-informed importance sampling
\citep{bowling2008strategy}, and baseline control variates
\citep{davidson2013baseline}. AIVAT combines action- and chance-informed
corrections to reduce variance while preserving the target expectation under its
randomization assumptions \citep{burch2018aivat}. It made shorter human-scale
DeepStack comparisons statistically informative \citep{moravvcik2017deepstack},
and related baseline control variates accelerate Monte Carlo game solving
\citep{schmid2019variance,davis2020low}.  A complementary route leaves the
estimator algebra untouched and instead correlates the sampling itself:
correlated chance sampling replaces i.i.d. chance draws in Monte Carlo CFR with
a persistent low-discrepancy stream per chance node
\citep{li2026correlatedchancesamplingmonte}.  Both lines reduce the noise of a
solving process; our target is the noise of the evaluation stream that compares
finished agents.

These methods principally improve an estimator within a specified evaluation
design. They do not, by themselves, validate stopping after repeated inspection,
nor do they state how a value function may be updated from an adaptively growing
evaluation stream. Our predictable interface fills this gap: the value
for hand $t$ may use earlier hands, while the resulting correction must remain
conditionally mean-zero under known randomization and must not depend on hand
$t$. Exact bounded-CS validity then also requires a genuine almost-sure bound on
the corrected outcome, which the game structure supplies for Leduc
(Section~\ref{sec:exp-leduc}).

\paragraph{Confidence sequences and anytime-valid inference.}
Confidence sequences arise from time-uniform concentration
\citep{howard2021time}, betting constructions
\citep{grunwald2024proposer}, time-uniform CLT theory and asymptotic
confidence sequences \citep{waudby2024time}, and the broader connection
between e-values, betting, and game-theoretic statistics
\citep{ramdas2023game}. Their defining advantage over repeatedly viewed
fixed-sample intervals is compatibility with continuous monitoring and optional
stopping under the assumptions of the chosen construction.  Inside
imperfect-information games, the same machinery has been used to decide when an
observed opponent tendency is established firmly enough to exploit, gating
restricted responses on a confidence sequence over pooled action frequencies and
committing only strategies whose exploitation certificate lies within a declared
budget \citep{li2026agentscertifyexploitsconfidencescheduled}.  There the
sequential interval licenses an action against an opponent; here it licenses a
verdict about which of two agents is stronger.

The two constructions we use sit at opposite ends of this literature. A bounded
betting-based CS gives exact finite-sample, time-uniform validity when the
stream has a constant conditional mean, the bets are predictable, and the
declared almost-sure bound is valid. A time-uniform CLT approximation instead
tracks the realized variance directly, and its guarantee is asymptotic: here the
AsympCS relies on a conditional Lindeberg condition and a nondegenerate limiting
averaged conditional variance. We therefore use the AsympCS as the efficient
primary interval and the bounded EB-CS as the form that supplies a finite-sample
certificate once an independent bound is available, and we report finite-run
coverage frequencies as the empirical evidence beside those guarantees.

Our stopping-cost analysis is specific to the construction. For the predictable
EB-CS whose bets are capped at $1/2$, the declared bound induces a deterministic
half-width floor proportional to $B/t$, so a variance gain becomes a
finite-target stopping gain only when that bound is tight relative to
$\sigma^2/\varepsilon$. This is what motivates the order-level design proxy of
Section~\ref{sec:arx-stopping-cost}.

\paragraph{Sources and provenance of online values.}
Modern game-playing systems routinely obtain values from learned or precomputed
components. DeepStack uses neural continuation values, and Libratus uses a
precomputed blueprint during real-time subgame solving
\citep{moravvcik2017deepstack,brown2018superhuman}. EVPA uses previously trained
expected-value networks for online pruning and abstraction
\citep{li2025efficient}, while Parallel CFR batches neural leaf evaluation in a
real-time CPU--GPU pipeline \citep{li2026real}. Student of Games learns such
values jointly for perfect- and imperfect-information play
\citep{schmid2023student}. These systems concern value
use in decision making or solving. For evaluation, the central issue is temporal
provenance: which observations were available when each correction value was
produced. Recording that provenance makes the no-current-hand restriction
checkable and distinguishes a legitimately updated value model from post hoc
rescoring.

\paragraph{Selective stopping and reproducibility.}
Poker evaluation is statistically fragile \citep{alon2007poker,dedonno2008poker},
and per-game costs create an incentive to inspect results early, but economic
motivation does not make selective stopping valid. Our release model targets one
evidentiary task: reproducing a stopping claim from the corrected outcomes
through the claimed stopping time and the complete stopping metadata, including
the bound and its provenance. This fixes the minimum release needed to recheck a
claim. A separate line of work studies whether the correction generator follows
its declared construction: post hoc AIVAT value-function tuning can induce
spurious significance \citep{kim2026heuristic}. Such estimator-integrity checks
complement the stopping-claim reconstruction studied here.

\section{Preliminaries}
\label{sec:arx-setup}

\subsection{Sequential agent evaluation}

An evaluator observes independent hands played by a fixed agent against a
fixed opponent profile.  Let $\F_{t-1}$ contain the complete trajectories of
hands $1{:}t-1$, all evaluator training and model-selection randomness revealed
before hand $t$, and all side information used to choose the current correction.
We assume this enlarged filtration does not reveal the current hand's random
actions.  Hand $t$ produces the evaluated agent's raw payoff
$X_t\in[-B_X,B_X]$, where $B_X$ is a declared structural bound, and
\[
  \E[X_t\mid\F_{t-1}]=\mu.
\]
Under the fixed-agent, fixed-opponent setup the raw hands are i.i.d.; we write
$\sigma_X^2=\Var(X_t)$. The target is a confidence interval for the unknown
expected payoff $\mu$ with half-width at most $\varepsilon$. A fixed-sample
interval does not license repeatedly inspecting the interval and stopping at a
data-dependent time. We therefore use confidence sequences.

A $(1-\alpha)$ confidence sequence (CS) is an adapted sequence
$\mathrm{CI}_t=[\widehat\mu_t\pm\hw_t]$ satisfying
\begin{equation}
\label{eq:arx-cs-guarantee}
 \Prob\!\left(\forall t\ge1:\ \mu\in\mathrm{CI}_t\right)\ge1-\alpha.
\end{equation}
This time-uniform event licenses optional stopping and continuation
\citep{howard2021time,ramdas2023game}. For target precision
$\varepsilon$, define the first crossing
\begin{equation}
\label{eq:arx-stopping-time}
 \stope(\varepsilon):=\inf\{t:\hw_t\le\varepsilon\}.
\end{equation}

\subsection{AIVAT corrections}
\label{sec:arx-aivat}

A correction at a node replaces the realized continuation value by its average
over the known action kernel, so two things must hold: the kernel is known, and
the decision to correct at that node is made before the action there is seen.
The notation below states both.

Let $H_c$ be a fixed finite set of eligible chance and evaluated-agent decision
nodes.  For hand $t$ and $h\in H_c$, let $\mathcal G^-_{t,h}$ denote the
sigma-field immediately before the action at $h$; it contains $\F_{t-1}$, the
within-hand history through reaching $h$, and the evaluated agent's internal
state already revealed in that history.  Let $S_{t,h}\in\{0,1\}$ be an optional
enablement indicator measurable with respect to $\mathcal G^-_{t,h}$.  The known
kernel $p_{t,h}$ must satisfy
\[
 \Prob(A_{t,h}=a\mid\mathcal G^-_{t,h})=p_{t,h}(a)\quad\text{a.s.}
\]
for every enabled node and action.  Thus enablement cannot depend on the
realized outgoing action or later information.  If $I_{t,h}$ indicates that $h$
is reached, define
\begin{equation}
\label{eq:arx-aivat}
\begin{aligned}
 Y_t&=X_t+C_t,\\
 C_t&=\sum_{h\in H_c}S_{t,h}I_{t,h}\left(
   \sum_a p_{t,h}(a)v_t(h\!\cdot\!a)-v_t(h\!\cdot\!A_{t,h})
 \right).
\end{aligned}
\end{equation}
AIVAT uses known action distributions only; unknown opponent decision nodes are
not correction points. Accurate continuation values can make $C_t$ cancel much
of the chance and evaluated-agent action variation in $X_t$
\citep{burch2018aivat,white2009learning}. Section~\ref{sec:arx-validity}
shows that $v_t$ need not be frozen before evaluation: predictability is the
relevant validity condition.

\subsection{The two confidence sequences}

Two constructions run side by side throughout the paper. The first is exact at
every look and pays for that exactness through the range it declares; the second
adapts to the realized variance and is valid asymptotically. AV-AIVAT reports
both.

The exact certificate is the predictable plug-in empirical-Bernstein CS
(EB-CS) studied here. For a stream declared to lie in $[-B,B]$, rescale
$Z_t=(Y_t+B)/(2B)\in[0,1]$. Let
$\widetilde m_{t-1}=(1/2+\sum_{s<t}Z_s)/t$ be the predictable regularized
plug-in mean, which is $\F_{t-1}$-measurable, and choose predictable bets
$\lambda_t\in[0,1/2]$. With
\[
 v_t^{\mathrm{EB}}=4(Z_t-\widetilde m_{t-1})^2,
 \qquad
 \psi_E(\lambda)=\frac{-\log(1-\lambda)-\lambda}{4},
\]
and $\loga=\log(2/\alpha)$, the unit-scale interval is centered at the
bet-weighted mean with the stated half-width,
\begin{equation}
\label{eq:arx-ebcs}
 \widehat m^{\mathrm{EB},[0,1]}_t=
 \frac{\sum_{s\le t}\lambda_s Z_s}{\sum_{s\le t}\lambda_s},
 \qquad
 \hw^{\mathrm{EB},[0,1]}_t=
 \frac{\loga+\sum_{s\le t}v_s^{\mathrm{EB}}\psi_E(\lambda_s)}
      {\sum_{s\le t}\lambda_s},
\end{equation}
intersected with $[0,1]$; when the denominator is zero the full interval
$[0,1]$ is used.

The payoff-scale interval applies the affine map
$m\mapsto2Bm-B$, so its center is $2B\,\widehat m^{\mathrm{EB},[0,1]}_t-B$ and
its half-width is $2B\,\hw^{\mathrm{EB},[0,1]}_t$. For bounded observations with
constant conditional mean and predictable bets, this is exactly the predictable
plug-in empirical-Bernstein confidence sequence
\citep{grunwald2024proposer}, which is time-uniform at level $1-\alpha$.
The bet cap $\lambda_t\le1/2$ is part of this construction, and
Section~\ref{sec:arx-stopping-cost} shows it is what turns the declared range
into a width floor.

Our practical primary interval is the asymptotic CS (AsympCS)
\citep{waudby2024time}. With running mean
$\widehat\mu_t=t^{-1}\sum_{s\le t}Y_s$, sample variance
$\widehat\sigma_t^2=(t-1)^{-1}\sum_{s\le t}(Y_s-\widehat\mu_t)^2$, and a fixed
tuning parameter $\rho>0$, it is centered at $\widehat\mu_t$ with half-width
\begin{equation}
\label{eq:arx-asympcs}
 \hw^{\mathrm A}_t=
 \widehat\sigma_t\,
 \sqrt{\frac{2(t\rho^2+1)}{t^2\rho^2}
 \log\!\left(\frac{\sqrt{t\rho^2+1}}{\alpha}\right)}.
\end{equation}
This is the standardized parameterization used in our implementation, in which
$\rho$ tunes the standardized process. Its guarantee is asymptotic in the sense
of \citet{waudby2024time}: under the martingale-difference regularity
stated in Proposition~\ref{prop:arx-asympcs-validity}, its endpoints are
almost surely asymptotically equivalent to those of an exact confidence
sequence.

\section{The AV-AIVAT Protocol}
\label{sec:arx-protocol}

Algorithm~\ref{alg:arx-protocol} states the full procedure.  AV-AIVAT assigns a
distinct role to each component: the CS makes continuous monitoring valid, the
predictable AIVAT interface permits past-only value updates, and AIVAT reduces
the payoff variance that can otherwise make anytime poker evaluation
impractical.

Three design choices are backed by the theory of
Section~\ref{sec:arx-theory}.
\emph{First}, the AsympCS is the primary interval and the EB-CS a conservative
certificate: by Theorem~\ref{thm:arx-eb-stopping} the EB-CS pays its
cap-induced floor in the $B\gg\sigma$ regime, while the AsympCS follows the
variance-ratio deterministic width benchmark asymptotically under its stated
regularity conditions.  Reporting both makes the trade-off explicit, and
Section~\ref{sec:arx-validity} fixes which of the two EB-CS modes a given
declared bound licenses.  AsympCS tuning is selected once on a separate
replication-level calibration split and frozen before evaluation, which rules
out post hoc retuning.
\emph{Second}, the value function may be refit online, on the very data being
evaluated, at any predictable schedule
(Proposition~\ref{prop:arx-predictable-validity});
Theorem~\ref{thm:arx-variance-regret} bounds the efficiency cost relative to an
oracle value function by the value-learner's variance regret.  This turns
AIVAT's most awkward practical requirement, a good $v$ \emph{before} any data
exist, into a warm start.
\emph{Third}, claim rechecking needs only the corrected prefix and stop
metadata: another evaluator recomputes both CSs at the claimant's stopping time
(Section~\ref{sec:auditing}), and a paired log may optionally be released so a
reader can inspect the raw stream.  Section~\ref{sec:release-protocol} gives the
full release protocol.

At hand $t$, the evaluator fixes $v_t$ using only information in
$\F_{t-1}$, plays and observes the hand, computes $Y_t$ by
\eqref{eq:arx-aivat}, updates both confidence sequences, and may then use the
new hand to construct $v_{t+1}$. At any data-dependent stop, the evaluator
reports both intervals, the corrected prefix $Y_{1:t}$, the stopping rule and
index, the declared corrected-stream bound $B_Y$, and all interval settings.
A paired $(X_t,Y_t)$ log may optionally be released so a reader can inspect the
raw stream; it is not required to recompute a claim from the corrected prefix.

AV-AIVAT separates three statements. The EB-CS is the exact bounded-stream
certificate. The AsympCS is the practical asymptotic interval under its stated
regularity conditions. Online refitting is valid only when every training,
tuning, and model-selection choice affecting hand $t$ is
$\F_{t-1}$-measurable; hand $t$ may update $v_{t+1}$, never $v_t$.

\section{Theory}
\label{sec:arx-theory}

\begin{algorithm}[t]
\caption{AV-AIVAT: anytime-valid variance-reduced evaluation}
\label{alg:arx-protocol}
\begin{algorithmic}[1]
\REQUIRE level $\alpha$, first eligible look $b$, locked CS settings, initial value function $v_1$; independently justified $B_Y$ for exact EB-CS mode
\FOR{$t = 1, 2, \dots$ (stop at any eligible time)}
  \STATE before each correction action, record its conditional kernel $p_{t,h}$ and choose $S_{t,h}$ without observing that action
  \STATE play hand $t$; observe payoff $X_t$ and trajectory $\omega_t$
  \STATE $Y_t \leftarrow X_t + \sum_{h \in H_c} S_{t,h} I_{t,h}
     \bigl( \sum_a p_{t,h}(a) v_t(h\!\cdot\!a) - v_t(h\!\cdot\!A_{t,h}) \bigr)$
  \STATE at $t\ge b$, update locked AsympCS \eqref{eq:arx-asympcs} on $Y_{1:t}$
     \COMMENT{asymptotic screen}
  \STATE if $B_Y$ is independently justified, update EB-CS \eqref{eq:arx-ebcs}
     \COMMENT{exact certificate}
  \STATE optionally refit $v_{t+1}$ on data through hand $t$
     \COMMENT{$v_t$ already fixed hand $t$}
\ENDFOR
\ENSURE report both applicable intervals at the data-dependent stopping time; publish $Y_{1:t}$, first-look/censoring metadata, kernel and enablement provenance, and optionally paired $(X_t,Y_t)$ diagnostics
\end{algorithmic}
\end{algorithm}

Three results support the protocol. The first licenses the predictable
plug-in interface, the second characterizes which confidence sequence to
report at a given target precision, and the third bounds what online value
learning costs relative to an oracle. Proofs are in
Appendices~\ref{app:arx-validity}--\ref{app:arx-online-proofs}. Throughout,
the evaluator declares a payoff bound used by both streams and $\loga$
abbreviates $\log(2/\alpha)$.

\subsection{Validity: Predictable Value Functions Suffice}
\label{sec:arx-validity}

The classical AIVAT analysis freezes its value function before the evaluation
sample is observed. In a sequential analysis, the weaker condition needed for
the correction argument is that the value function used on the current hand is
fixed before that hand is observed.

\begin{proposition}[Exact validity under predictable value functions]
\label{prop:arx-predictable-validity}
Let $\F_t=\sigma(\text{hands }1{:}t)$ and suppose that each $v_t$ is
$\F_{t-1}$-measurable with $\|v_t\|_\infty\le V$. At every correction
node $h$, assume that $p_h$ is the conditional action kernel: given
$\F_{t-1}$, the full history through reaching $h$, and any evaluated-agent
internal state included in that history, the realized action has law $p_h$.
Assume
$\E[X_t\mid\F_{t-1}]=\mu$ and that at most $K$ correction points occur on a
hand. For $Y_t$ defined by \eqref{eq:arx-aivat}:
\begin{enumerate}
\item $\E[C_t\mid\F_{t-1}]=0$, and therefore
      $\E[Y_t\mid\F_{t-1}]=\mu$;
\item $|Y_t|\le B_X+2KV=:B_Y$;
\item the EB-CS \eqref{eq:arx-ebcs}, rescaled using $B_Y$, satisfies
      \eqref{eq:arx-cs-guarantee} exactly.
\end{enumerate}
\end{proposition}

\begin{proposition}[Asymptotic validity of the AsympCS]
\label{prop:arx-asympcs-validity}
Assume the conditions of Proposition~\ref{prop:arx-predictable-validity} and
let $V_t=\sum_{s\le t}\Var(Y_s\mid\F_{s-1})$. If $V_t\to\infty$ almost surely,
the sample variance is relatively consistent
($t\,\widehat\sigma_t^2/V_t\to1$ a.s.), $t^{-1}V_t$ converges to a finite
nonzero limit, and the martingale differences satisfy the conditional
Lindeberg condition, then the AsympCS \eqref{eq:arx-asympcs} is an asymptotic
confidence sequence in the sense of \citet{waudby2024time}.
\end{proposition}

\paragraph{Proof mechanism.}
Let $H_c$ be the fixed finite set of chance and evaluated-agent correction
nodes in one hand. The random trajectory sum can be written
\begin{equation}
\label{eq:arx-nodewise-correction}
 C_t=\sum_{h\in H_c}\mathbf1\{h\text{ reached in hand }t\}
 \left(\E_{a\sim p_h}[v_t(h\!\cdot\!a)]
       -v_t(h\!\cdot\!a_h)\right).
\end{equation}
Conditioning on $\F_{t-1}$ fixes $v_t$. Conditional also on reaching $h$, the
realized action $a_h$ has the known law $p_h$, so the parenthesized term has
conditional mean zero. The reach event is determined before the action at
$h$; multiplying by its indicator and summing the finite node set therefore
gives $\E[C_t\mid\F_{t-1}]=0$. This argument does not require the opponent's
strategy because opponent nodes are excluded from $H_c$.

Each correction summand has absolute value at most $2V$, proving the stated
bound. After rescaling by $B_Y$, the stream is bounded in $[0,1]$ and has a
constant conditional mean. These are exactly the inputs used by the published
predictable plug-in EB-CS theorem. For intuition, a predictable capital factor
$1+\lambda_t(Z_t-\mu_Z)$ has conditional expectation one at the true
$\mu_Z$; the full EB-CS claim relies on the cited e-process theorem rather than
on an identification with this simpler factor.
Proposition~\ref{prop:arx-asympcs-validity} rests on a different set of
hypotheses, the martingale-difference Lindeberg and averaged
conditional-variance conditions, which is why the two streams carry separate
guarantees throughout the paper. Appendix~\ref{app:arx-validity} gives both
proofs in full.

The interface these propositions license is an aggressive one: the value
function may be refit after every observed hand, provided every consequence
for hand $t$ is measurable with respect to $\F_{t-1}$. That measurability
requirement is what makes the refitting safe, since it keeps hand $t$ out of
its own correction, out of any post hoc choice among corrections, and out of
the choice of the declared bound $B_Y$. Fitting the value function on the same
observations to which its corrections are applied is precisely the
nonpredictable failure mode excluded here \citep{kim2026heuristic}.

\subsection{Which Confidence Sequence? A Stopping-Time
Characterization}
\label{sec:arx-stopping-cost}

We next isolate why a large variance reduction need not produce a comparable
stopping-time reduction for the particular EB-CS in
\eqref{eq:arx-ebcs}. Throughout this section, $\loga=\log(2/\alpha)$.

\begin{lemma}[Bet-cap width floor]
\label{lem:arx-bet-floor}
For every data realization and every predictable bet sequence
$\lambda_s\in[0,1/2]$, the payoff-scale EB-CS on a stream declared in
$[-B,B]$ satisfies
\[
 \hw_t^{\mathrm{EB}}\ge \frac{4B\loga}{t}.
\]
\end{lemma}

\paragraph{Mechanism.}
All terms in the numerator of \eqref{eq:arx-ebcs} are nonnegative, while
$\sum_{s\le t}\lambda_s\le t/2$. Thus the unit-scale half-width is at least
$2\loga/t$; mapping a unit-scale width back to a range of length $2B$ gives the
claim. This is a deterministic, variance-independent floor for this capped-bet
construction.

\begin{theorem}[Stopping-time characterization for the studied EB-CS]
\label{thm:arx-eb-stopping}
Fix $\alpha\in(0,1/2)$ and an i.i.d. stream supported on $[-B,B]$ with
constant variance $\sigma^2>0$.
\begin{enumerate}
\item For every predictable bet sequence,
\[
 \stope_{\mathrm{EB}}(\varepsilon)\ge
 \frac{4B\loga}{\varepsilon}.
\]
\item For every $\delta\in(0,1)$ and every $\varepsilon$ below a
threshold depending on the distribution and $\delta$, there is a
constant bet $\lambda_\varepsilon$, depending on $(\varepsilon,\sigma,B)$,
such that, with probability at least $1-\delta$,
\[
 \stope_{\mathrm{EB}}(\varepsilon)\le
 8\max\!\left\{\frac{B\loga}{\varepsilon},
                 \frac{\sigma^2\loga}{\varepsilon^2}\right\}.
\]
\item If $\varepsilon\le\sigma/(2\sqrt3)$ and
$a:=\sqrt{\sigma^2+4\varepsilon^2}\le B$, there are two i.i.d. laws on the
common support $\{-a,a\}$, both with variance $\sigma^2$ and with means
$\pm2\varepsilon$, such that every valid CS stopping at half-width
$\varepsilon$ satisfies, for at least one law,
\[
 \E[\stope(\varepsilon)]\ge
 \frac{(1-2\alpha)\log((1-\alpha)/\alpha)}
 {(2\varepsilon/a)\log((a+2\varepsilon)/(a-2\varepsilon))}.
\]
For fixed $\alpha<1/2$, this is
$\Omega(\sigma^2\log(1/\alpha)/\varepsilon^2)$.
\end{enumerate}
\end{theorem}

\paragraph{Proof mechanisms.}
Part 1 is Lemma~\ref{lem:arx-bet-floor}. For Part 2, work on $[0,1]$ and use
$\psi_E(\lambda)\le\lambda^2/4$. With a constant bet and
$V_t=\sum_{s\le t}v_s^{\mathrm{EB}}$,
\[
 \hw_t^{\mathrm{EB},[0,1]}
 \le \frac{\loga}{\lambda t}+\frac{\lambda V_t}{4t}.
\]
Bounded i.i.d. concentration yields $V_t\le8t\sigma_Z^2$ eventually and
uniformly with the required probability. The oracle choice
$\lambda=\min\{1/2,\varepsilon_Z/(4\sigma_Z^2)\}$ then separates the
range-controlled $1/\varepsilon$ regime from the variance-controlled
$1/\varepsilon^2$ regime. This is an existence result for a distribution-aware
constant bet, not a guarantee for every adaptive betting rule.

For Part 3, tilt a common Rademacher support $\{-a,a\}$ to means
$\pm2\varepsilon$. A width-$\varepsilon$ confidence interval covering the
positive mean has positive center, while one covering the negative mean has
negative center. Time-uniform coverage therefore induces a sequential test with
both errors at most $\alpha$. Its information per sample is exactly
$(2\varepsilon/a)\log((a+2\varepsilon)/(a-2\varepsilon))$, giving the stated
expected lower bound. Full details are in
Appendix~\ref{app:arx-stopping-proofs}.

\begin{remark}[The design proxy]
\label{rem:arx-design-proxy}
The deterministic floor, oracle-bet high-probability upper bound, and
common-support expected lower-bound family motivate the design proxy
\begin{equation}
\label{eq:arx-general-proxy}
 \frac{\max\{B_X,\sigma_X^2/\varepsilon\}}
      {\max\{B_Y,\sigma_Y^2/\varepsilon\}}.
\end{equation}
It captures the order of magnitude at stake for a design choice rather than a
pathwise, expected, or distribution-wise stopping-time equivalence. If
$B_X=B_Y=B$ and $0<\sigma_Y^2\le\sigma_X^2$, elementary case analysis rewrites
this proxy as
\begin{equation}
\label{eq:arx-clip-proxy}
 \clip\!\left(\frac{\sigma_X^2}{B\varepsilon},
              1,\frac{\sigma_X^2}{\sigma_Y^2}\right),
\end{equation}
which reads off three regimes: range-limited, neutral, and variance-limited. If
AIVAT corrections require $B_Y>B_X$, the general expression
\eqref{eq:arx-general-proxy} applies.
\end{remark}

\begin{proposition}[Deterministic AsympCS width benchmark]
\label{prop:arx-asymp-stopping}
Write the AsympCS half-width \eqref{eq:arx-asympcs} as
$\hw^{\mathrm A}_t=\widehat\sigma_t\,g(t)$, where
$g(t)=\sqrt{2(t\rho^2+1)t^{-2}\rho^{-2}\log(\sqrt{t\rho^2+1}/\alpha)}$ depends
only on $t,\rho,\alpha$ and is independent of the stream. For each fixed
$\sigma>0$, define the deterministic asymptotic width benchmark
\[
 t_{\varepsilon}(\sigma):=
 \inf\{t\ge t_0:\sigma g(t)\le\varepsilon\},
\]
where $t_0$ is any deterministic index after which $g$ is decreasing. Then
$t_{\varepsilon}(\sigma)\to\infty$ as $\varepsilon\to0$, and for two positive
scales $\sigma_X\ge\sigma_Y$,
\[
 \frac{t_{\varepsilon}(\sigma_X)}{t_{\varepsilon}(\sigma_Y)}
 =\frac{\sigma_X^2}{\sigma_Y^2}\,L_\varepsilon,
\]
where $L_\varepsilon$ is the ratio of the corresponding slowly varying
logarithmic factors. In particular, this deterministic benchmark has the usual
variance-ratio scaling up to that logarithmic correction.
\end{proposition}

The benchmark here is the solution of a deterministic asymptotic width equation,
which is the object the AV-AIVAT design turns on: AsympCS serves as the efficient
primary interval because its width tracks the realized variance, and the studied
EB-CS serves as the exact finite-sample certificate whose speed is set by the
declared range and bet cap. How much of a variance gain that certificate can
convert into earlier stopping is then a question about how tight the declared
bound is, which Section~\ref{sec:exp-leduc} answers for a game whose structure
supplies an analytic one.

\subsection{Online Value Functions: a Variance-Regret Bound}
\label{sec:arx-online-values}

Proposition~\ref{prop:arx-predictable-validity} establishes that predictable
online value functions preserve the relevant conditional mean. We now state the
efficiency counterpart: how much a value function learned during the evaluation
costs relative to one that was right from the start. The statement is made
against an explicit asymptotic width model, which the AsympCS satisfies.

Let
$\sigma_s^2:=\Var(Y_s\mid\F_{s-1})$, where the conditional variance depends
on $v_s$. Let $\vstar$ be a comparator value function with per-hand variance
$\sstar^2>0$, and define
$\overline\sigma_t^2=t^{-1}\sum_{s\le t}\sigma_s^2$.

\begin{assumption}[Variance-adaptive width]
\label{ass:arx-variance-width}
Uniformly over deterministic $u\in[1,2]$, as $t\to\infty$,
\[
 \hw_{\lfloor ut\rfloor}^2
 =\frac{\overline\sigma_{\lfloor ut\rfloor}^2
         g_{\lfloor ut\rfloor}}{\lfloor ut\rfloor}
   (1+o_{\Prob}(1)),
 \qquad
 \frac{g_{\lfloor ut\rfloor}}{g_t}\to1
\]
uniformly, where the remainder is understood as
$\sup_{u\in[1,2]}|\cdot|\xrightarrow{\Prob}0$ and $g_t>0$ is deterministic and
slowly varying. The deterministic oracle asymptotic benchmark
$\stope^\star(\varepsilon)$ is defined as the asymptotic solution of
\begin{equation}
\label{eq:arx-oracle-benchmark}
 \varepsilon^2=\frac{\sstar^2g_t}{t}.
\end{equation}
For AsympCS, $g_t$ retains its logarithmic factor.
\end{assumption}

In words, the squared half-width tracks the average conditional variance
accumulated so far, divided by the number of hands and inflated by the slowly
varying factor $g_t$; the benchmark $\stope^\star(\varepsilon)$ is the horizon at
which that expression reaches $\varepsilon^2$ when every hand is scored by the
comparator $\vstar$. It is therefore a deterministic quantity fixed by
$\sstar^2$, $g_t$, and $\varepsilon$, and plays the same role here that the
deterministic width benchmark $t_\varepsilon(\sigma)$ of
Proposition~\ref{prop:arx-asymp-stopping} plays for a fixed variance scale, with
the running average $\overline\sigma_t^2$ in place of a constant $\sigma^2$. In
the notation of Section~\ref{sec:arx-stopping-cost} the two factors are the same
object, $g_t=t\,g(t)^2$, written per-hand here because the assumption is stated
for the squared half-width.

\begin{theorem}[Variance regret gives a one-sided no-delay guarantee]
\label{thm:arx-variance-regret}
Under Assumption~\ref{ass:arx-variance-width}, suppose
\[
 \sum_{s\le t}(\sigma_s^2-\sstar^2)\le R_t,
\]
where $R_t$ is a deterministic, nonnegative, nondecreasing sequence with
$R_t=o(t)$. Then, as $\varepsilon\to0$, with probability tending to one,
\begin{equation}
\label{eq:arx-regret-transfer}
 \stope(\varepsilon)
 \le \stope^\star(\varepsilon)
 \left(
  1+\frac{2R_{2\stope^\star(\varepsilon)}}
           {\sstar^2\stope^\star(\varepsilon)}+o(1)
 \right).
\end{equation}
\end{theorem}

\paragraph{Proof mechanism.}
Write $\tau^\star=\stope^\star(\varepsilon)$ and
$\delta=2R_{2\tau^\star}/(\sstar^2\tau^\star)$. Sublinear regret makes
$\delta\le1$ eventually. At the deterministic candidate time
$t_\delta=\lceil(1+\delta)\tau^\star\rceil\le2\tau^\star$, monotonicity of
$R_t$ gives
\[
 \overline\sigma_{t_\delta}^2
 \le\sstar^2\left(1+\frac{\delta}{2(1+\delta)}\right).
\]
Combining this inequality with \eqref{eq:arx-oracle-benchmark}, the uniform
width expansion, and slow variation of $g_t$ yields
\[
 \frac{\hw_{t_\delta}^2}{\varepsilon^2}
 \le \frac{2+3\delta}{2(1+\delta)^2}(1+o_{\Prob}(1)).
\]
The deterministic factor lies below one by at least $\delta/8$ for
$\delta\in(0,1]$. A vanishing enlargement handles the case in which $\delta$
is smaller than the stochastic remainder, proving
\eqref{eq:arx-regret-transfer}. Appendix~\ref{app:arx-online-proofs} supplies
the complete argument.

The next lemma connects value-function approximation to variance regret.

\begin{lemma}[Value-function perturbation]
\label{lem:arx-value-perturbation}
If $\|v-\vstar\|_\infty\le\Delta$, then pathwise
\[
 |Y_t(v)-Y_t(\vstar)|\le2K\Delta,
\]
and therefore
\[
 \sigma^2(v)-\sstar^2\le4K\sstar\Delta+4K^2\Delta^2.
\]
\end{lemma}

The pathwise statement follows by subtracting the two AIVAT corrections: each
of at most $K$ summands changes by at most $2\Delta$. The variance statement
then follows from the triangle inequality in conditional $L^2$.

\begin{corollary}[Polynomial value learning has asymptotically vanishing delay]
\label{cor:arx-polynomial-rate}
If $\|v_t-\vstar\|_\infty\le c t^{-\beta}$ for some
$\beta\in(0,1]$, then one may take
\[
 R_t=O\!\left(K\sstar c\,S_\beta(t)+K^2c^2S_{2\beta}(t)\right),
 \qquad S_q(t):=\sum_{s\le t}s^{-q}=o(t),
\]
and hence, with probability tending to one,
\[
 \stope(\varepsilon)\le
 \stope^\star(\varepsilon)(1+o(1)).
\]
\end{corollary}

The guarantee is one-sided, which is the direction an evaluator needs: it caps
any asymptotic delay relative to the deterministic oracle width benchmark, so a
value learner whose variance regret is sublinear reaches a target precision at
the oracle's own asymptotic horizon, and it does so while the exact EB-CS
validity of Section~\ref{sec:arx-validity} continues to hold. Learning during
the run therefore costs nothing asymptotically, and the cost at finite horizons
is what Section~\ref{sec:exp-e2} measures.
Appendix~\ref{app:arx-online-proofs} records what a matching lower bound would
additionally require.

\begin{remark}[Rarely visited information sets]
\label{rem:arx-rare-visits}
The exponent $\beta$ of Corollary~\ref{cor:arx-polynomial-rate} is governed by
the \emph{least-visited} information set rather than by the number of hands.
For a Laplace-smoothed opponent frequency model,
$\|v_t-\vstar\|_\infty\asymp\sqrt{\log t/N_{\min}(t)}$ with $N_{\min}(t)$ the
minimum visit count. Reach probabilities in Leduc span two orders of magnitude,
so $N_{\min}(t)\ll t$ and the effective $\beta$ is small. The residual
finite-horizon gap to oracle in Section~\ref{sec:exp-e2} is what this predicts.
\end{remark}

\section{Experiments}
\label{sec:experiments}

We evaluate AV-AIVAT on three sources. P0 is a HUNL evaluation
corpus containing 15 PokerSkill/LLM agent configurations and paired per-hand
raw/AIVAT outcomes from fixed-opponent multi-hand runs. Requiring both fields
leaves 4,028--5,000 hands per run and 71,439 total. E1 and E2 use Leduc
hold'em, where exact full-tree evaluation supplies ground truth.  A2 uses the
HUNL payoff pool for a simulated rolling leaderboard that monitors continuously
and stops at the first favorable look.  All analyses use nominal $\alpha=0.05$;
experiment-specific seeds and AsympCS settings are recorded in the appendix.
E1 and E2 followed a frozen internal protocol; P0 used an approved fixed-seed
gate without an independently timestamped preregistration; A2 and the later
sensitivity checks are labeled accordingly.

\paragraph{Targeted implementation regression test.}
With an oracle value function and corrections at every node in the controlled
Leduc laboratory, AIVAT must telescope to $Y_t\equiv\mu$ hand by hand.  The
targeted regression test satisfies $\max_t|Y_t-\mu|=9.1\times10^{-16}$ over 200
hands. This test guards the implementation of the telescoping identity.

\subsection{HUNL evaluation}
\label{sec:exp-hunl}

AIVAT sharply reduces empirical variance.  Across the 15 runs, the median raw
to AIVAT variance ratio is $54.4\times$ (range $24.2$--$86.0\times$).  At a
$\pm1$ BB target, the median ratio of raw to AIVAT AsympCS stopping times is
$74.17\times$ (range $54.53$--$97.62\times$), whereas the corresponding EB-CS
ratio is $1.365\times$ (range $1.235$--$1.556\times$) over 14 evaluable runs;
one run's raw stream failed to cross in most reshufflings.  A stopping-time
ratio is a nonlinear path statistic and need not equal the variance ratio.

At the tighter $\pm0.5$ BB target, all 15 raw AsympCS streams are censored at
their available run horizons (4,028--5,000 hands).  The AIVAT AsympCS median stopping time, computed within
each run over its 200 reshufflings and then summarized across the 15 run-level
medians, is 56.5 hands, with range 42.5--71.

\begin{table}[t]
\centering
\small
\begin{tabular}{@{}lc@{}}
\toprule
Quantity (15 HUNL runs; median [range]) & Value \\
\midrule
Variance ratio $\sigma_X^2/\sigma_Y^2$ & $54.4\times$ [$24.2$, $86.0$] \\
AsympCS stop ratio, $\pm1$ BB & $74.17\times$ [$54.53$, $97.62$] \\
EB-CS stop ratio, $\pm1$ BB ($n=14$)$^\dagger$ & $1.365\times$ [$1.235$, $1.556$] \\
Raw AsympCS, $\pm0.5$ BB & 15/15 censored \\
AIVAT AsympCS, $\pm0.5$ BB (hands) & 56.5 [42.5, 71] \\
\bottomrule
\end{tabular}
\caption{P0 stopping and variance summaries.  Each stopping entry first takes
the median over 200 reshufflings within a run and then summarizes run-level
values.  The EB-CS ratio has 14 evaluable runs; the other rows use all 15.
$^\dagger$The HUNL EB-CS row declares $B=200$ BB, which the 200 BB stacks bound
structurally on the raw stream but not on the corrected one, so the ratio is a
descriptive replay of the boundary.}
\label{tab:hunl}
\end{table}

\begin{figure}[t]
\centering
\includegraphics[width=\linewidth]{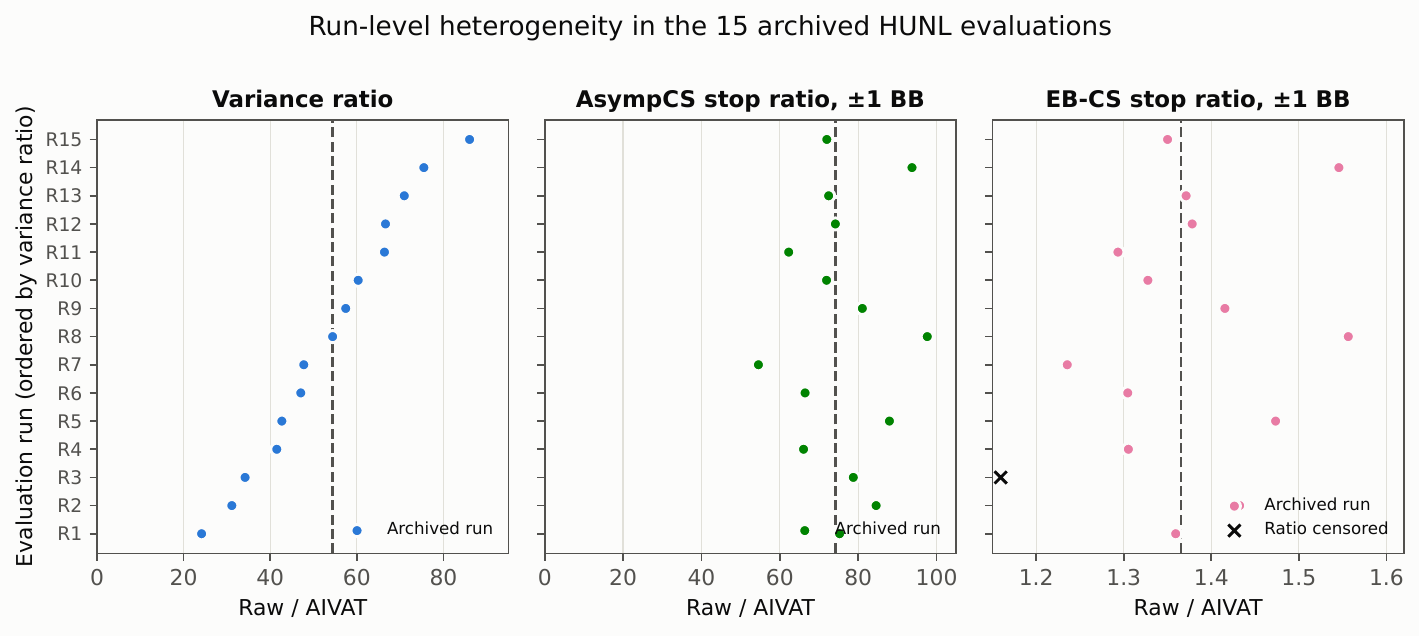}
\caption{Run-level heterogeneity in the 15 HUNL evaluations.  Each
point is one configuration.  Dashed lines mark cross-run medians, and the
cross in the EB panel denotes the one ratio censored because the raw stream
failed to cross in most reshufflings.}
\label{fig:hunl-forest}
\end{figure}

\paragraph{Bound provenance and finite-horizon calibration.}
P0 declares $B=200$ BB for all 15 logs.  On the raw stream this value is
structural rather than data-chosen: the evaluation seats 200 BB stacks and
forbids rebuys, so a single hand transfers at most one effective stack and
$|X_t|\le200$ BB holds almost surely.  The corpus is consistent with that
ceiling, attaining it exactly and never exceeding it.  The corrected stream has
no matching analytic ceiling here, because an AIVAT correction accumulates over
transitions and scales with the value estimates rather than with the stack, so
exact certification of the AIVAT arm would require a separate almost-sure
argument, which is what the Leduc structural certificate supplies in
Section~\ref{sec:exp-leduc}.  The reported EB-CS stopping ratio therefore pairs
an exact raw boundary with a descriptive corrected one.  The
AsympCS false-positive summary reported by P0 is 7.1067\%, the unweighted mean
of 30 condition-level rates, namely raw and AIVAT for each of 15 runs with
1,000 resamples per condition; the AIVAT-only mean is 12.43\%.  The
corresponding EB-CS summary is 0.01\%, which is small but not zero, so the two
boundaries separate on this corpus exactly as the width floor predicts.  Removing any one
run and recomputing the medians moves the variance ratio only within
$[51.1,55.9]$ and the two stopping ratios within $[73.3,74.7]$ and
$[1.359,1.371]$, so no single configuration drives the contrast
(Appendix~\ref{app:robustness}).

\subsection{Leduc ground-truth experiments}
\label{sec:exp-leduc}

The hero is CFR$^+$ after 2,000 iterations and the opponent is CFR$^+$ after 20
iterations.  Exact evaluation gives $\mu=-0.0047697$ chips per hand.  E1 uses
400 replications of 4,000 hands.  The empirical standard deviations are 3.96
(raw) and 1.39 (AIVAT), a variance ratio of $8.08\times$.  Median AsympCS
stopping ratios are $5.29\times$ at $\pm0.5$ chips and $7.91\times$ at
$\pm0.2$; at $\pm0.1$, every raw stream is censored while AIVAT stops at median
1,789 hands.

A frozen structural rerun makes the E1 EB-CS coverage claim exact.  Standard
Leduc has $|X|\le13$ and $|v|\le13$; corrected deal, chance, and hero
transitions telescope, leaving at most four opponent transitions, so
$|Y|\le117$.  The complete legal tree verifies the transition counts and the
telescoping identity, and Appendix~\ref{app:experimental-details} gives the
derivation. Across
400 replications, EB-CS excludes the exact mean in 2 raw and 0 AIVAT streams;
AsympCS rates are 3.75\% and 2.25\%.  Exactness here comes from the theorem
together with the structural bound, and these counts are the finite-horizon
Monte Carlo evidence beside it.

The same structural bound lets us read off the price of exactness.  Under
$B_Y=117$, empirical-to-floor ratios are 1.0019 and 1.0074 at $t=1000$ and
$4000$; the loose $B'=200$ comparison gives 1.0006 and 1.0025.  The certificate
runs within about 1\% of the deterministic floor its declared range induces,
which is why the floor formula, and not the variance ratio, predicts how fast an
exact interval can be.  The earlier sample-derived corrected bound 22 and its
descriptive execution are recorded in
Appendix~\ref{app:experimental-details}.

\begin{figure}[t]
\centering
\includegraphics[width=0.9\linewidth]{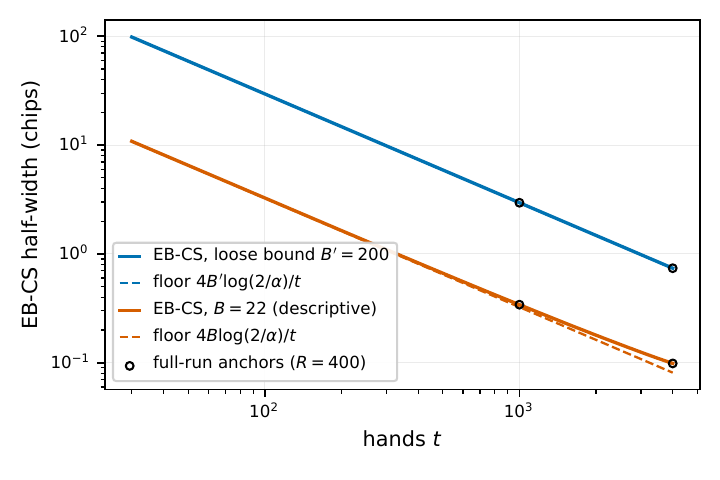}
\caption{EB-CS width curves on Leduc under different declared bounds.  The loose
$B'=200$ curve nearly equals its deterministic floor $4B'\log(2/\alpha)/t$; the
sample-derived $B=22$ curve is descriptive.  The analytic
$B_Y=117$ floor anchors are reported in the text.}
\label{fig:arx-floor}
\end{figure}

\subsection{Controlled comparison of predictable online values}
\label{sec:exp-e2}

E2 compares raw, frozen, predictable, and oracle arms on the same 100 sets of
8,192 simulated hands. The predictable arm refits a Laplace-smoothed opponent
action-frequency model at doubling epochs using past hands only, which isolates
the learning mechanism under specified opponent-model mismatches.

\begin{table}[t]
\centering
\small
\begin{tabular}{@{}lrrrr@{}}
\toprule
 & raw & frozen & predictable & oracle \\
\midrule
Std. (chips) & 3.97 & 1.80 & 1.49 & 1.40 \\
$\stope(\pm0.5)$ & 607 & 164.5 & 156 & 118.5 \\
$\stope(\pm0.2)$ & 3759.5 & 754 & 649 & 473 \\
$\stope(\pm0.1)$ & cens. & 3078 & 2280 & 1786 \\
\bottomrule
\end{tabular}
\caption{E2 median AsympCS stopping times on paired simulated hands.}
\label{tab:e2}
\end{table}

The primary profile gives the aggregate summary in Table~\ref{tab:e2}; at
$\pm0.1$, predictable stops 25.9\% earlier than frozen while oracle remains
faster. A three-profile extension clarifies the boundary. In a matched-uniform
control, frozen and oracle are identical and no oracle gap is identifiable. In
two opponent-model mismatch profiles, predictable learning recovers 78.6\% and
77.1\% of the frozen-to-oracle variance gap. At $\pm0.1$, predictable medians
are 2240.5 and 2608 hands, versus 3039 and 4337.5 for frozen and 1744 and 1425.5
for oracle; raw streams are fully right-censored. These finite-horizon results
quantify partial adaptation under both tested mismatches.

\begin{figure}[t]
\centering
\includegraphics[width=0.9\linewidth]{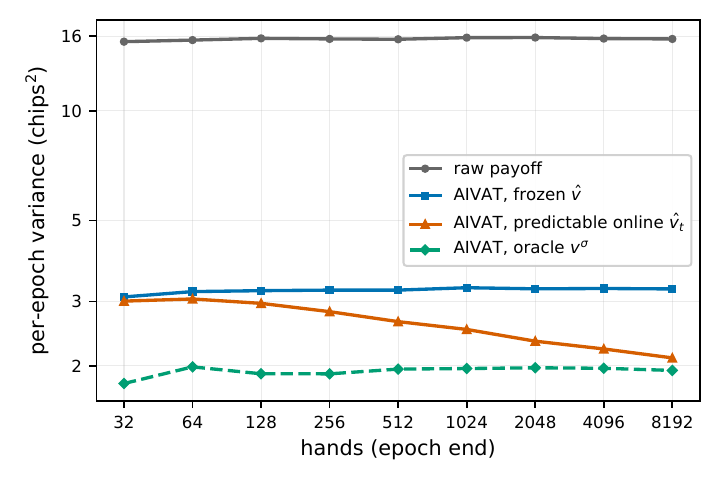}
\caption{E2 per-epoch variance under the controlled construction. Predictable
past-only learning moves toward the oracle over the observed horizon, while the
frozen and raw baselines remain flat.}
\label{fig:arx-variance}
\end{figure}

\paragraph{Held-out AsympCS calibration.}
A separate protocol uses 400 calibration and 400 evaluation replications with
disjoint seeds. The 18-candidate grid varies internal $\alpha$, $t_{\rm opt}$,
and burn-in, then freezes a single candidate before generating evaluation data.
Held-out exclusion counts are raw 8/400, frozen 5/400, predictable
7/400, and oracle 5/400. The largest four-arm Bonferroni-adjusted one-sided
Clopper--Pearson upper bound is 4.21\%, so the registered empirical gate passes.

\section{Continuous Monitoring and Time-Uniform Validity}
\label{sec:auditing}

AV-AIVAT (Section~\ref{sec:arx-protocol}) lets an evaluator recompute a CS on
the exact corrected prefix behind a published early-stopping claim and check it
\emph{at the reported stopping time}.  This needs the corrected prefix and stop
metadata, but no new games.  The simulation below exercises that recheck: it
measures how often an ordinary fixed-sample interval manufactures significance
under continuous monitoring, and whether the two prespecified sequential
boundaries retain the claim at the selected prefix.

\subsection{Optional stopping under continuous monitoring}
\label{sec:audit-a2a}

We form a zero-effect pool by centering the AIVAT payoffs within each of the 15
HUNL runs and concatenating the resulting 71,439 values.  Each of 2,000
simulated leaderboard entries samples 5,000 hands with replacement.  A monitor
inspects a naive fixed-sample 95\% interval after every hand from $t=30$ and
stops at the first exclusion of zero.  We then recompute EB-CS and AsympCS on
the same selected prefix and check the interval at that stopping time.

Stopping at the first favorable look claims a nonzero effect in 1,227 of 2,000
entries (61.35\%), with median stopping time 59 hands among claims, even though
the true effect is zero by construction.  Checking a time-uniform interval on
that same selected prefix screens those claims out: EB-CS contains zero at all
1,227 selected stops, and AsympCS at 1,226 of them (99.9185\%).  Run as
standalone monitors over the full 5,000 hands, EB-CS has observed false-positive
rate 0\% and AsympCS 10.4\%.

Each simulated entry declares its own $B$ as the ceiling of that entry's
observed maximum absolute payoff, so the EB column
replays the exact boundary descriptively on the resampled pool, and the AsympCS
column is the asymptotic screen whose finite-horizon calibration is visible in
its standalone rate.  What the comparison isolates is the effect of prefix
selection, which is common to both columns.

\begin{table}[t]
\centering
\small
\begin{tabular}{@{}lc@{}}
\toprule
A2 $H_0$ quantity & Result \\
\midrule
Fixed-CI claims at first favorable look & 1,227/2,000 (61.35\%) \\
Median claimed stop & 59 hands \\
EB-CS contains zero at selected stop & 1,227/1,227 \\
AsympCS contains zero at selected stop & 1,226/1,227 \\
Standalone EB-CS detection by 5,000 & 0/2,000 \\
Standalone AsympCS detection by 5,000 & 208/2,000 \\
\bottomrule
\end{tabular}
\caption{A2 continuous-monitoring simulation under a constructed zero-effect
pool.  Continuous monitoring of a fixed-sample interval yields a 61\%
false-positive rate; checking either CS on the selected prefix keeps zero inside
the interval and so screens the claim out.  The EB entries declare per-entry
observed bounds, so they replay the boundary descriptively.}
\label{tab:a2-h0}
\end{table}

\paragraph{Genuine effects are still recovered.}
Adding constant shifts of 0.05, 0.1, and 0.2 BB per hand gives claim rates of
86.6\%, 97.4\%, and 100\% at the first favorable look.  Confirmation at that same
selected stop remains 0\% for the descriptive EB analysis and 0.23\%, 1.44\%,
and 7.0\% for AsympCS.  In contrast, when AsympCS continues as its own monitor,
it detects the 0.1 effect in 81.2\% of entries with median stopping time 960.5
and the 0.2 effect in 100\% with median 173.  The distinction matters: rejecting
an unsupported early claim is not the same as denying the underlying effect,
which can be detected after additional hands by a sequential monitor.

\begin{figure}[t]
\centering
\includegraphics[width=0.78\linewidth]{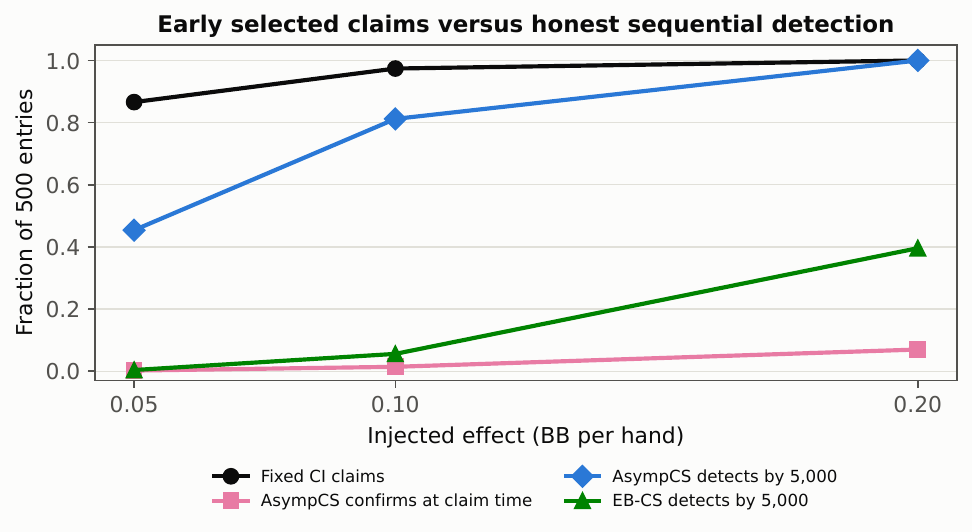}
\caption{Injected-effect experiment.  A first-crossing fixed CI often claims
success before a sequential boundary crosses.  Continuing AsympCS under its own
stopping rule detects the larger injected effects by 5,000 hands.}
\label{fig:a2-power}
\end{figure}

\section{Release Protocol}
\label{sec:release-protocol}

An AV-AIVAT claim is auditable only if its statistical inputs and stopping
decision can be reconstructed.  We therefore recommend releasing one record
per evaluated hand together with immutable run-level metadata.

\paragraph{Required outcome stream.}
For each hand, release its sequential index, the corrected payoff $Y_t$, units,
and an identifier linking the record to one evaluation run.  Release the raw
payoff $X_t$ when available.  The corrected prefix is sufficient to recompute a
CS; releasing paired $(X_t,Y_t)$ additionally lets a reader inspect the raw
stream.  Missing or invalid rows must be marked rather than silently
renumbered, and the rule that maps platform records to analyzed records must be
stated.

\paragraph{Correction provenance.}
Identify the correction algorithm, the known action distributions used at
correction nodes, and the value-function artifact or training procedure.  For
online values, record update epochs and enough state to verify that $v_t$ was
fixed before observing hand $t$.  If a value function is frozen, report what
data trained it and confirm that evaluation hands were not used to fit it.

\paragraph{Inference specification.}
Release the confidence-sequence implementation and version, confidence level,
burn-in, target half-width, rescaling convention, bet cap, censoring rule, and
all numerical constants.  A bounded exact-coverage claim must include an
independently justified almost-sure bound for the corrected payoff.  Rules of the
game supply such bounds directly, through a maximum per-hand contribution or a
seated stack that no hand can exceed.  A bound chosen instead from the analyzed
observations, including an observed maximum rounded upward, must be labeled
data-dependent and the resulting EB analysis descriptive rather than exact.  Approximate AsympCS results must be labeled
asymptotic and accompanied by finite-horizon calibration where available.

\paragraph{Stopping metadata.}
Release the declared stopping rule, target quantity, first eligible look,
reported stopping index $\hat t$, whether the run was censored, and the exact
prefix used for the claim.  If analysis occurred after a fixed horizon rather
than online, say so.  An auditor should be able to recompute the interval at
$\hat t$ without collecting new data or guessing which rows were included.

\paragraph{Experimental provenance.}
Release random seeds, sample sizes, configuration identifiers, code commit or
archive hash, and machine-readable result files.  Distinguish preregistered,
internally frozen, exploratory, and post hoc analyses.  Deviations must be
recorded next to the affected claim.  In this study that includes E1's change
from the registered raw-payoff bound $B=13$ to the observed-joint-maximum-based
$B=22$, as well as the observed-bound choices in P0, E2 coverage, and A2.

\paragraph{Minimal recheck procedure.}
Given the released prefix, another evaluator (i) verifies ordering and row
selection; (ii) reconstructs the stated correction and predictability
assumptions where artifacts permit; (iii) recomputes the declared interval at
every eligible look and at $\hat t$; and (iv) checks an independently justified
corrected-payoff bound if exact EB validity is claimed.  This reconstructs the
stopping claim.  Detecting faults or adversarial changes in the correction
generator is a complementary estimator-audit task.

\section{Discussion}
\label{sec:discussion}

Within the operating scope of Appendix~\ref{app:scope}, the two AV-AIVAT streams
divide the labor cleanly.  The AsympCS is the working instrument: P0's median variance
factor of $54.4$ becomes a median $74\times$ reduction in the hands needed to
reach $\pm1$ BB, and its finite-horizon exclusion rates at nominal 5\% are
7.1067\% averaged over the 30 P0 raw/AIVAT run-stream conditions and 10.4\%
under continuous monitoring through 5,000 hands in A2.  These rates are the
calibration evidence for an asymptotic screen, and the held-out gate of
Section~\ref{sec:exp-e2} was frozen in advance so that the evidence is
prospective rather than selected after the fact.

The EB-CS is the certificate, and what it certifies depends on where the
declared bound comes from.  The structural Leduc rerun declares the analytic
$B_Y=117$ and therefore instantiates the exact-validity theorem directly.  The
remaining executions declare bounds derived from observed maxima, so their
EB-CS coverage summaries replay an execution descriptively;
Appendix~\ref{app:experimental-details} records the provenance of each declared
bound alongside the corresponding result.  Separating these two modes is what
lets one paper report an exact certificate and a practical screen without
conflating them.

For practice, the stopping ratio matters because hands are expensive.  At the
\$0.07--\$0.30 per evaluated hand reported for LLM agents in full poker games, a
$74\times$ reduction in the hands required for a target precision turns a budget
question into a routine one, and the reduction costs nothing in auditability:
the corrected prefix together with stop metadata is enough for a third party to
recheck the interval at the claimant's own stopping time, which is exactly the
recheck A2 performs against an invalid naive monitor.

The width floor identifies the sharpest open problem.  A bet cap forces
$\hw^{\mathrm{EB}}_t\ge4B\log(2/\alpha)/t$, so an exact interval converts a
variance gain into earlier stopping only when the declared bound is tight
relative to $\sigma^2/\varepsilon$.  Leduc shows this is achievable: the
analytic certificate follows from the telescoping structure of the game tree,
and the resulting EB-CS runs within about 1\% of its floor at both measured
horizons.  An analytic corrected-payoff bound for HUNL, or a confidence sequence
whose range dependence adapts to the realized payoff scale, would extend exact
certification to the setting where evaluation is most expensive.
Appendix~\ref{app:future-work} collects the further directions this opens.

\section{Conclusion}
\label{sec:conclusion}

Past-only value functions preserve the mean-zero correction law under the
required action kernel and pre-action enablement, so the value model may keep
learning while an evaluation runs instead of being frozen before it starts.  The
AsympCS turns the resulting variance reduction into much earlier stopping, and
the EB-CS certifies a stopping claim exactly once a corrected-payoff bound is
established independently of the evaluation sample, as the game structure
permits in Leduc.  A bet-capped exact interval pays a width floor set by that
declared bound, which is what decides how much of a variance gain becomes
earlier stopping.  Releasing the corrected prefix, stop metadata, and bound
provenance makes the resulting claim recomputable at the claimant's own stopping
time.  AV-AIVAT therefore specifies adaptation, stopping, and bound provenance
together.

\bibliographystyle{plainnat}
\bibliography{references}

\appendix
\section{Proofs of Predictable Validity}
\label{app:arx-validity}

We prove Propositions~\ref{prop:arx-predictable-validity}
and~\ref{prop:arx-asympcs-validity}. Fix the finite game
tree for one hand. For $h\in H_c$ let $I_{t,h}=\mathbf1\{h\text{ reached}\}$ be
the reach indicator, $A_{t,h}$ the realized action, and $S_{t,h}$ the
enablement indicator. Let $\mathcal G^-_{t,h}$ be the sigma-field immediately
before the action at $h$; it contains $\F_{t-1}$ and the within-hand history
through reaching $h$. Then \eqref{eq:arx-aivat} equals the fixed-node sum
\begin{equation}
\label{eq:arx-app-fixed-node-sum}
 C_t=\sum_{h\in H_c}S_{t,h}I_{t,h}
 \left(\sum_a p_{t,h}(a)v_t(h\!\cdot\!a)
       -v_t(h\!\cdot\!A_{t,h})\right).
\end{equation}
Writing the correction over a fixed finite set avoids conditioning on the
random set of visited correction points.

\subsection{Conditional mean}

Fix $h\in H_c$. The factors $S_{t,h}$, $I_{t,h}$, $v_t$, and the kernel
$p_{t,h}$ are all $\mathcal G^-_{t,h}$-measurable: $v_t$ is
$\F_{t-1}$-measurable, and the reach and enablement events are determined
before the action at $h$. By the conditional-kernel hypothesis, given
$\mathcal G^-_{t,h}$ the realized action $A_{t,h}$ has law $p_{t,h}$, so
\[
 \E\!\left[v_t(h\!\cdot\!A_{t,h})\mid\mathcal G^-_{t,h}\right]
 =\sum_a p_{t,h}(a)\,v_t(h\!\cdot\!a).
\]
Hence each node term has $\mathcal G^-_{t,h}$-conditional mean zero. Taking
$\E[\,\cdot\mid\F_{t-1}]$ and using the tower property
($\F_{t-1}\subseteq\mathcal G^-_{t,h}$) makes each summand's
$\F_{t-1}$-conditional mean zero, without conditioning on a possibly null reach
event. Summing the finite set $H_c$ in \eqref{eq:arx-app-fixed-node-sum} gives
$\E[C_t\mid\F_{t-1}]=0$. Since $\E[X_t\mid\F_{t-1}]=\mu$ by hypothesis,
\[
 \E[Y_t\mid\F_{t-1}]=\mu.
\]
The unknown opponent strategy can affect whether a correction node is reached,
but not the known action law at such a node. This is why no correction is made
at an unknown opponent decision node.

\subsection{Boundedness}

For every reached correction node,
\[
 \left|\E_{a\sim p_h}[v_t(h\!\cdot\!a)]
       -v_t(h\!\cdot\!a_h)\right|
 \le 2\|v_t\|_\infty\le2V.
\]
At most $K$ correction nodes occur on a realized trajectory, so
$|C_t|\le2KV$. Combining this with $|X_t|\le B_X$ gives
\[
 |Y_t|\le B_X+2KV=B_Y.
\]

\subsection{Exact EB-CS validity}

Rescale $Z_t=(Y_t+B_Y)/(2B_Y)\in[0,1]$ and define
$\mu_Z=(\mu+B_Y)/(2B_Y)$. The preceding conditional-mean calculation gives
$\E[Z_t\mid\F_{t-1}]=\mu_Z$.

For intuition, for any predictable $\lambda_t\in[0,1/2]$,
\[
 M_t^+(\mu_Z)=\prod_{s\le t}\bigl(1+\lambda_s(Z_s-\mu_Z)\bigr)
\]
is a nonnegative unit-initialized martingale because
\begin{align*}
 \E[M_t^+(\mu_Z)\mid\F_{t-1}]
 &=M_{t-1}^+(\mu_Z)
   \left(1+\lambda_t\E[Z_t-\mu_Z\mid\F_{t-1}]\right)\\
 &=M_{t-1}^+(\mu_Z).
\end{align*}
The analogous process with the minus sign is also a nonnegative martingale.
This calculation identifies the role of predictability and the constant
conditional mean.

The exact interval in \eqref{eq:arx-ebcs} is the predictable plug-in
empirical-Bernstein construction \citep{grunwald2024proposer}. Its
published e-process theorem applies to bounded adapted observations with
constant conditional mean and predictable bets. The preceding parts establish
those hypotheses. Inverting that e-process therefore yields
\[
 \Prob\!\left(\forall t\ge1:\mu_Z\in
 \mathrm{CI}_t^{\mathrm{EB},[0,1]}\right)\ge1-\alpha.
\]
The affine map from $[0,1]$ back to $[-B_Y,B_Y]$ preserves coverage, proving
\eqref{eq:arx-cs-guarantee} for the corrected payoff mean and completing the
proof of Proposition~\ref{prop:arx-predictable-validity}. The
empirical-Bernstein boundary comes from the published e-process result; the
simpler capital process above is explanatory and is not used to derive
\eqref{eq:arx-ebcs}.

\subsection{Asymptotic AsympCS validity}

The process $D_t=Y_t-\mu$ is an adapted martingale-difference sequence with
$|D_t|\le2B_Y$. Write $\sigma_t^2=\Var(Y_t\mid\F_{t-1})$ and
$V_t=\sum_{s\le t}\sigma_s^2$. We discharge, in turn, the hypotheses of the
martingale asymptotic confidence sequence of \citet{waudby2024time}.

First, $t^{-1}V_t\to\sigma_\infty^2\in(0,\infty)$ and the monotonicity of $V_t$
give $V_t\to\infty$ almost surely. Second, since $|D_t|\le2B_Y$, for any fixed
$\kappa\in(0,1)$ the truncation level $V_t^{\kappa}$ eventually exceeds
$(2B_Y)^2$, so the conditional Lindeberg tail terms vanish for all large $t$ and
their series converges almost surely. Third, the relative-consistency
requirement $t\,\widehat\sigma_t^2/V_t\to1$ a.s. follows from the martingale
strong law: with $Q_t=\sum_{s\le t}D_s^2$ and $S_t=\sum_{s\le t}D_s$, the
bounded-increment martingale $Q_t-V_t$ has conditional second moment $O(V_t)$,
so $(Q_t-V_t)/V_t\to0$ a.s., and $S_t^2/(tV_t)\to0$ a.s.; since
$t\,\widehat\sigma_t^2/V_t=Q_t/V_t-S_t^2/(tV_t)$, the ratio tends to one.
Boundedness alone does not imply the positive limiting variance, which is why it
is assumed. Under these three conditions the cited result supplies the
asymptotic guarantee for \eqref{eq:arx-asympcs} in its stated sense, completing
the proof of Proposition~\ref{prop:arx-asympcs-validity}.

\section{Proofs for the Stopping-Cost Results}
\label{app:arx-stopping-proofs}

\subsection{Proof of Lemma~\ref{lem:arx-bet-floor}}

Since $v_s^{\mathrm{EB}}\ge0$ and $\psi_E(\lambda_s)\ge0$, Equation
\eqref{eq:arx-ebcs} gives
\[
 \hw_t^{\mathrm{EB},[0,1]}
 \ge\frac{\loga}{\sum_{s\le t}\lambda_s}
 \ge\frac{\loga}{t/2}=\frac{2\loga}{t}.
\]
Mapping a unit-scale half-width back to $[-B,B]$ multiplies it by $2B$.
Therefore $\hw_t^{\mathrm{EB}}\ge4B\loga/t$ pointwise for every realization
and predictable bet sequence.

\subsection{Proof of Theorem~\ref{thm:arx-eb-stopping}}

\paragraph{Part 1.}
If $\hw_t^{\mathrm{EB}}\le\varepsilon$, Lemma~\ref{lem:arx-bet-floor}
forces $t\ge4B\loga/\varepsilon$.

\paragraph{Part 2: rescaling and the boundary.}
Work on $[0,1]$ with
\[
 \varepsilon_Z=\frac{\varepsilon}{2B},\qquad
 \sigma_Z^2=\frac{\sigma^2}{4B^2}.
\]
For $\lambda\in(0,1/2]$,
\begin{align*}
 -\log(1-\lambda)-\lambda
 &=\sum_{k\ge2}\frac{\lambda^k}{k}\\
 &\le\frac{\lambda^2}{2}\sum_{j\ge0}\lambda^j
 =\frac{\lambda^2}{2(1-\lambda)}\le\lambda^2.
\end{align*}
Hence $\psi_E(\lambda)\le\lambda^2/4$. For a constant bet $\lambda$ and
$V_t=\sum_{s\le t}v_s^{\mathrm{EB}}$,
\begin{equation}
\label{eq:arx-app-constant-bet-width}
 \hw_t^{\mathrm{EB},[0,1]}
 \le\frac{\loga}{\lambda t}+\frac{\lambda}{4}\frac{V_t}{t}.
\end{equation}

\paragraph{Part 2: controlling the empirical-variance proxy.}
Because $Z_s\in[0,1]$,
$v_s^{\mathrm{EB}}=4(Z_s-\widehat\mu_{s-1})^2\in[0,4]$, and
\begin{equation}
\label{eq:arx-app-proxy-conditional-mean}
 \E[v_s^{\mathrm{EB}}\mid\F_{s-1}]
 =4\left(\sigma_Z^2+(\widehat\mu_{s-1}-\mu_Z)^2\right).
\end{equation}
For the bounded i.i.d. stream, sample-mean concentration and a union bound over
time imply that, for every $\delta\in(0,1)$, with probability at least
$1-\delta/2$ there is a deterministic
$t_1=t_1(\delta,\sigma_Z)$ such that
\[
 \sum_{s\le t}\E[v_s^{\mathrm{EB}}\mid\F_{s-1}]
 \le6t\sigma_Z^2\quad\text{for all }t\ge t_1.
\]
The process obtained by subtracting the predictable sum in
\eqref{eq:arx-app-proxy-conditional-mean} from $V_t$ is a martingale with
increments in $[-4,4]$. A time-uniform Hoeffding bound
\citep{howard2021time} implies, on an event of probability at least
$1-\delta/2$,
\[
 V_t\le\sum_{s\le t}\E[v_s^{\mathrm{EB}}\mid\F_{s-1}]
 +c\sqrt{t\log(\log t/\delta)}
\]
simultaneously in $t$, for a universal constant $c$, for all $t$ beyond the
fixed first eligible look. Combining the two events,
there is a deterministic $t_0=t_0(\delta,\sigma_Z)$ such that, with probability
at least $1-\delta$,
\begin{equation}
\label{eq:arx-app-proxy-concentration}
 V_t\le8t\sigma_Z^2\quad\text{for all }t\ge t_0.
\end{equation}

\paragraph{Part 2: oracle constant bet.}
For the target $\varepsilon_Z$, choose
\[
 \lambda=\min\left\{\frac12,\frac{\varepsilon_Z}{4\sigma_Z^2}\right\}.
\]
This bet depends on the distributional variance but not on the observed data,
so it is predictable. On the event
\eqref{eq:arx-app-proxy-concentration}, the second term of
\eqref{eq:arx-app-constant-bet-width} is at most
$2\lambda\sigma_Z^2\le\varepsilon_Z/2$. The first term is at most
$\varepsilon_Z/2$ whenever
\[
 t\ge
 \begin{cases}
 4\loga/\varepsilon_Z,&\lambda=1/2,\\[2pt]
 8\sigma_Z^2\loga/\varepsilon_Z^2,
   &\lambda=\varepsilon_Z/(4\sigma_Z^2).
 \end{cases}
\]
Therefore
\[
 \stope_{\mathrm{EB}}(\varepsilon_Z)
 \le\max\left\{\frac{4\loga}{\varepsilon_Z},
                  \frac{8\sigma_Z^2\loga}{\varepsilon_Z^2}\right\}
 \vee t_0.
\]
Substitution of the payoff-scale quantities yields
\[
 \stope_{\mathrm{EB}}(\varepsilon)
 \le\max\left\{\frac{8B\loga}{\varepsilon},
                  \frac{8\sigma^2\loga}{\varepsilon^2}\right\}
 \vee t_0.
\]
For every $\varepsilon$ below the distribution- and $\delta$-dependent
threshold at which the maximum exceeds $t_0$, this proves Part 2.

\paragraph{Part 3: common-support alternatives.}
Let $a=\sqrt{\sigma^2+4\varepsilon^2}$ and define laws $P_+$ and $P_-$ on
$\{-a,a\}$ by
\[
 P_+(a)=\frac{1+2\varepsilon/a}{2},\qquad
 P_-(a)=\frac{1-2\varepsilon/a}{2},
\]
with complementary masses at $-a$. Their means are $2\varepsilon$ and
$-2\varepsilon$, and both variances equal
$a^2-4\varepsilon^2=\sigma^2$. The condition
$\varepsilon\le\sigma/(2\sqrt3)$ implies $2\varepsilon/a\le1/2$, so all
masses are positive; $a\le B$ places the common support inside $[-B,B]$.

If either expected stopping time is infinite, the claimed lower bound is
immediate. Otherwise, at the stopping time decide $P_+$ when the reported
interval center is positive and $P_-$ otherwise. A half-width-$\varepsilon$
interval that covers $2\varepsilon$ has positive center, and one that covers
$-2\varepsilon$ has negative center. Time-uniform coverage bounds both testing
errors by $\alpha$. The one-sample divergences are equal:
\[
 \mathrm{KL}(P_+\|P_-)=\mathrm{KL}(P_-\|P_+)
 =\frac{2\varepsilon}{a}
   \log\frac{a+2\varepsilon}{a-2\varepsilon}.
\]
The sequential information inequality therefore gives, for at least one law,
\[
 \E[\stope(\varepsilon)]\ge
 \frac{(1-2\alpha)\log((1-\alpha)/\alpha)}
 {(2\varepsilon/a)\log((a+2\varepsilon)/(a-2\varepsilon))}.
\]
For $x\in[0,1/2]$,
$x\log((1+x)/(1-x))\le(8/3)x^2$. Taking
$x=2\varepsilon/a$, the denominator is at most
$32\varepsilon^2/(3a^2)\le32\varepsilon^2/(3\sigma^2)$, which proves the
stated order for fixed $\alpha<1/2$.

\subsection{Derivation and boundary of the design proxy}

The theorem's three components have different quantifiers: a deterministic
construction-specific floor, an oracle-bet high-probability upper bound, and an
expected lower bound for a particular two-law family. They motivate
\eqref{eq:arx-general-proxy} but do not combine into a stopping-time
equivalence. When $B_X=B_Y=B$ and
$0<\sigma_Y^2\le\sigma_X^2$, consider the locations of
$\varepsilon$ relative to $\sigma_X^2/B$ and $\sigma_Y^2/B$. Direct case
analysis gives the clipped expression in \eqref{eq:arx-clip-proxy}. This
algebra establishes only the equivalence of the two proxy formulas.

\subsection{Proof of Proposition~\ref{prop:arx-asymp-stopping}}

Write $g(t)^2=t^{-1}\ell(t)$, where
\[
 \ell(t)=2\left(1+\frac{1}{t\rho^2}\right)
 \log\!\left(\frac{\sqrt{t\rho^2+1}}{\alpha}\right)
\]
is positive and slowly varying. For sufficiently large deterministic $t_0$,
$g$ is decreasing to zero, so
$t_\varepsilon(\sigma)=\inf\{t\ge t_0:\sigma g(t)\le\varepsilon\}$ is
well-defined and diverges as $\varepsilon\to0$. At the asymptotic solution,
\[
 t_\varepsilon(\sigma)
 =\frac{\sigma^2}{\varepsilon^2}
   \ell\!\left(t_\varepsilon(\sigma)\right)(1+o(1)),
\]
where integer rounding is asymptotically immaterial. Applying this identity to
$\sigma_X$ and $\sigma_Y$ gives
\[
 \frac{t_\varepsilon(\sigma_X)}{t_\varepsilon(\sigma_Y)}
 =\frac{\sigma_X^2}{\sigma_Y^2}
  \frac{\ell(t_\varepsilon(\sigma_X))}
       {\ell(t_\varepsilon(\sigma_Y))}(1+o(1)).
\]
The final quotient is precisely the ratio of slowly varying logarithmic factors
denoted by $L_\varepsilon$. This calculation compares deterministic width
benchmarks only. It does not use, or imply, monotonicity of the sample variance
or of the realized AsympCS half-width, and therefore does not characterize an
actual random first crossing.

\section{Proofs for Online Value Functions}
\label{app:arx-online-proofs}

\subsection{Proof of Theorem~\ref{thm:arx-variance-regret}}

Write $\tau^\star=\stope^\star(\varepsilon)$ and define
\[
 \delta_\varepsilon=
 \frac{2R_{2\tau^\star}}{\sstar^2\tau^\star},
 \qquad
 t_\delta=\left\lceil(1+\delta_\varepsilon)\tau^\star\right\rceil.
\]
Because $R_t=o(t)$ and $\tau^\star\to\infty$ as
$\varepsilon\to0$, eventually
$R_{2\tau^\star}\le\sstar^2\tau^\star/2$. Hence
$\delta_\varepsilon\le1$ and $t_\delta\le2\tau^\star$ because
$\tau^\star$ is represented by an integer benchmark time. Monotonicity of
$R_t$ and the variance-regret bound yield
\begin{align}
 \overline\sigma_{t_\delta}^2
 &\le\sstar^2+\frac{R_{t_\delta}}{t_\delta}\notag\\
 &\le\sstar^2+
 \frac{R_{2\tau^\star}}{(1+\delta_\varepsilon)\tau^\star}
 +o((\tau^\star)^{-1})\notag\\
 &=\sstar^2\left(
 1+\frac{\delta_\varepsilon}{2(1+\delta_\varepsilon)}
 \right)+o((\tau^\star)^{-1}).
\label{eq:arx-app-average-variance}
\end{align}

By definition of the deterministic oracle asymptotic benchmark,
\[
 \varepsilon^2=\frac{\sstar^2g_{\tau^\star}}{\tau^\star}.
\]
This equality uses an asymptotically immaterial integer representative.
Assumption~\ref{ass:arx-variance-width}, uniformly on the deterministic window
$[\tau^\star,2\tau^\star]$, together with
\eqref{eq:arx-app-average-variance}, gives
\begin{align}
 \frac{\hw_{t_\delta}^2}{\varepsilon^2}
 &\le
 \frac{1+\frac{\delta_\varepsilon}{2(1+\delta_\varepsilon)}}
      {1+\delta_\varepsilon}
 \frac{g_{t_\delta}}{g_{\tau^\star}}
 (1+o_{\Prob}(1))\notag\\
 &=\frac{2+3\delta_\varepsilon}
         {2(1+\delta_\varepsilon)^2}
 (1+o_{\Prob}(1)).
\label{eq:arx-app-width-ratio}
\end{align}
For every $\delta\in(0,1]$,
\begin{equation}
\label{eq:arx-app-deterministic-gap}
 1-\frac{2+3\delta}{2(1+\delta)^2}
 =\frac{\delta+2\delta^2}{2(1+\delta)^2}
 \ge\frac{\delta}{8}.
\end{equation}

It remains to account for a possible $\delta_\varepsilon$ smaller than the
stochastic remainder. By uniform convergence in probability, choose a
deterministic sequence $\eta_\varepsilon\downarrow0$ such that the absolute
remainder in \eqref{eq:arx-app-width-ratio} is
$o_{\Prob}(\eta_\varepsilon)$. If
$\delta_\varepsilon\ge\eta_\varepsilon$, then
\eqref{eq:arx-app-deterministic-gap} implies
$\hw_{t_\delta}\le\varepsilon$ with probability tending to one. Therefore
\[
 \stope(\varepsilon)
 \le(1+\delta_\varepsilon)\tau^\star+1
\]
with probability tending to one. If
$\delta_\varepsilon<\eta_\varepsilon$, repeat the candidate-time argument with
$\eta_\varepsilon$ in place of $\delta_\varepsilon$; monotonicity of $R_t$
and $\delta_\varepsilon<\eta_\varepsilon$ preserve the required average-
variance upper bound, while the deterministic gap dominates the remainder.
Then
\[
 \stope(\varepsilon)
 \le(1+\eta_\varepsilon)\tau^\star+1.
\]
This inequality holds on an event whose probability tends to one. Combining the two cases and
using $1/\tau^\star=o(1)$ yields
\[
 \stope(\varepsilon)
 \le\stope^\star(\varepsilon)
 \left(
 1+\frac{2R_{2\stope^\star(\varepsilon)}}
          {\sstar^2\stope^\star(\varepsilon)}+o(1)
 \right)
\]
with probability tending to one.

\subsection{Proof of Lemma~\ref{lem:arx-value-perturbation}}

Set $d=v-\vstar$, so $\|d\|_\infty\le\Delta$. Subtracting the two corrected
estimators on the same realized hand gives
\[
 Y_t(v)-Y_t(\vstar)
 =\sum_{k\in K_t}\left(
 \E_{a\sim p_k}[d(h_k\!\cdot\!a)]-d(h_k\!\cdot\!a_k)
 \right).
\]
Every summand has absolute value at most $2\Delta$, and $|K_t|\le K$.
Therefore
\[
 |Y_t(v)-Y_t(\vstar)|\le2K\Delta.
\]
Let $D=Y_t(v)-Y_t(\vstar)$. The triangle inequality in conditional $L^2$
gives
\begin{align*}
 \sigma(v)
 &\le\sstar+\sqrt{\Var(D\mid\F_{t-1})}\\
 &\le\sstar+\|D\|_\infty
 \le\sstar+2K\Delta.
\end{align*}
Thus
\begin{align*}
 \sigma^2(v)-\sstar^2
 &=(\sigma(v)-\sstar)(\sigma(v)+\sstar)\\
 &\le(2K\Delta)(2\sstar+2K\Delta)\\
 &=4K\sstar\Delta+4K^2\Delta^2.
\end{align*}

\subsection{Proof of Corollary~\ref{cor:arx-polynomial-rate}}

If $\|v_t-\vstar\|_\infty\le ct^{-\beta}$, Lemma
\ref{lem:arx-value-perturbation} gives
\[
 \sigma_t^2-\sstar^2
 \le4K\sstar c\,t^{-\beta}+4K^2c^2t^{-2\beta}.
\]
Summing through $t$ shows that the regret condition holds with
\[
 R_t\le4K\sstar c\,S_\beta(t)+4K^2c^2S_{2\beta}(t),
 \qquad S_q(t)=\sum_{s\le t}s^{-q}.
\]
For every $q>0$, $S_q(t)=o(t)$: it grows as $O(t^{1-q})$ for $q<1$,
$O(\log t)$ for $q=1$, and $O(1)$ for $q>1$. Hence $R_t=o(t)$. Applying
Theorem~\ref{thm:arx-variance-regret} gives the one-sided statement
\[
 \stope(\varepsilon)\le
 \stope^\star(\varepsilon)(1+o(1))
\]
with probability tending to one. This does not imply convergence of the ratio
to one: the online procedure may cross earlier. A matching lower bound would
require additional control before the oracle benchmark time. The comparison is
to the deterministic benchmark from Assumption~\ref{ass:arx-variance-width},
not to an oracle sample path's random first crossing.

\section{Experimental Details}
\label{app:experimental-details}

\subsection{Analysis status and common settings}

E1 and E2 follow the hypotheses, sample sizes, and kill criteria of an internal
protocol frozen before those
experiments.  P0 used an internally approved fixed-seed gate with a
prespecified kill criterion, but no independently timestamped P0
preregistration is archived.  A2 is a subsequent fixed-seed audit application.
The leave-one-run-out and entry-length analyses are post hoc sensitivity
checks.  P0, the original E1/E2 executions, A2, and entry-length sensitivity
share one fixed seed and the default AsympCS setting $t_{\mathrm opt}=1000$ with
a 30-hand burn-in.  The cross-setting and held-out analyses use disjoint seed
ranges and the locked AsympCS settings reported with those analyses.  All experiments use
nominal $\alpha=0.05$.  The default AsympCS parameterization is
$\rho^2=2\log(1/\alpha)/t_{\mathrm opt}$; the EB-CS uses the regularized
predictable plug-in mean $\widetilde m_{t-1}$ and predictable bets capped at
$1/2$.

\subsection{HUNL data and P0 computation}

The HUNL corpus consists of 15 PokerSkill/LLM-agent configurations, each played
against GTO Wizard in one multi-hand run.  A configuration fixes its agent and
PokerSkill settings; the 15 configurations need not correspond to 15 distinct
base models.  One record per hand supplies the raw and the corrected payoff.
Requiring both leaves 4,028--5,000
paired hands per run and 71,439 hands overall.  Payoffs and means are expressed
in big blinds (BB) and BB per hand, respectively.  The
Figure~\ref{fig:hunl-forest} labels R1--R15 order the runs by variance ratio.

For each run, P0 creates 200 random permutations of the observed paired hands.
For each method, stream, and target $\varepsilon\in\{2,1,0.5,0.3,0.2\}$ BB,
it records the median stopping time and censoring fraction over permutations.
The AsympCS burn-in is 30 hands.  Run-level summaries in the main text take the
median and range across the 15 within-run medians; they do not pool all
permutations across runs.

At $\varepsilon=0.5$ BB, the 15 within-run AIVAT AsympCS median stopping times
are
\[
42.5,43.5,48,49,49,53,56,56.5,57,57,58,58.5,61,63,71.
\]
Thus the cross-run median is 56.5 and the range is 42.5--71.  Every raw stream
has censoring fraction one at that target.  The previously used range 53--388
mixed a different collection of fields and is not the reported estimand.

\paragraph{P0 bound provenance.}
The execution obtained 200 BB by taking the maximum absolute value over the
observed raw and AIVAT payoffs and rounding upward to a multiple of 50 BB, which
yields the same number for every P0 run.  On the raw stream the value is also
available structurally: the evaluation seats 200 BB stacks and forbids rebuys, so
one hand transfers at most one effective stack and $|X_t|\le200$ BB holds almost
surely for future draws as well.  Across the 71,439 analyzed hands the maximum
absolute raw payoff is exactly 200 BB, attained on 237 hands, with none
exceeding it.  The corrected stream is different: an AIVAT correction accumulates
over transitions and scales with the value estimates rather than with the stack,
so no analogous ceiling is available and the largest observed absolute corrected
payoff is 153.79 BB.  P0's AIVAT EB-CS summaries therefore remain descriptive
unless an independent $|Y_t|\le200$ BB argument is supplied.

\paragraph{P0 false-positive aggregation.}
For each of 15 runs and each of its raw and AIVAT streams, P0 centers the stream,
draws 1,000 bootstrap sequences with replacement, and monitors for an exclusion
of zero.  The stored aggregate AsympCS value 0.0710667 is the unweighted mean
of these 30 condition-level false-positive rates.  The AIVAT-only mean across
15 run conditions is 0.124267; its median is 0.100.  None of these is a single
false-positive probability estimated from one pooled resampling experiment.

\subsection{Leduc E1 and E2}

The Leduc testbed has three ranks, two suits, two betting rounds, bet sizes 2
and 4, and at most two raises per round.  The hero is CFR$^+$ after 2,000
iterations and the opponent is CFR$^+$ after 20 iterations.  Exact full-tree
evaluation gives $\mu=-0.00476973597$ chips per hand.  E1 uses 400 replications
of 4,000 hands.  E2 uses 100 replications of 8,192 hands and evaluates all four
arms on the same simulated hands.

\paragraph{E1 bound deviation.}
The frozen protocol specified the structural raw-payoff bound
$|X_t|\le13$ chips.  Corrected AIVAT samples exceeded this bound.  The execution
instead took
the observed maximum absolute value across all simulated raw and corrected
samples and rounded it upward to $B=22$.  The observed 0\% EB-CS miscoverage at
this executed bound is therefore descriptive, not a confirmatory validation of
exact coverage.  The deliberately loose $B'=200$ arm was unchanged and is used
only to demonstrate the numerical width floor.

\paragraph{E2 predictability and bound provenance.}
The frozen arm uses a uniform-opponent value function.  The predictable arm
refits a Laplace-smoothed opponent action-frequency model at doubling epochs
using hands strictly before the current hand.  The oracle arm uses the true
opponent continuation values.  Its EB coverage calculation takes the maximum
absolute observed predictable-arm sample
and rounds it upward.  Hence its reported 0\% EB-CS miscoverage is descriptive
and is not an exact finite-sample coverage certificate.  The AsympCS result is
also not exact: its guarantee is asymptotic and its observed miscoverage is
6.0\% over 100 replications.

\paragraph{Targeted zero-variance regression test.}
In the controlled laboratory only, using the oracle value function and
corrections at all nodes, including opponent nodes whose strategy is known,
causes AIVAT to telescope to $Y_t=\mu$ hand by hand.  The targeted regression
test has maximum absolute floating-point error $9.1\times10^{-16}$ over 200
hands. It guards the implementation of this identity; it does not validate a
learned value function or a real-world opponent-node correction.

\subsection{Structural-bound and calibration analyses}

The new structural E1 rerun uses analytic bounds $B_X=13$ and $B_Y=117$, a fixed
seed, 400 replications, and 4,000 hands.  The corrected bound is obtained as
follows.  In standard Leduc the pot never exceeds 13 chips, so $|X|\le13$ and
every continuation value satisfies $|v|\le13$.  The default estimator corrects
the initial deal, the community chance node, and every hero transition.  Along a
realized trajectory the corrected terms telescope: the value entering a
corrected node cancels against the value leaving the preceding corrected node,
so the sum collapses to the root value plus the transitions that are left
uncorrected.  Enumerating the complete legal tree shows at most four such
opponent transitions on any trajectory, and each contributes at most $2\cdot13$
in absolute value, giving $|Y|\le13+4(2\cdot13)=117$.  The enumeration verifies
both the transition counts and the telescoping identity.  This bound is
strategy-independent and uses no evaluation sample, so it is admissible as the
$B_Y$ of Proposition~\ref{prop:arx-predictable-validity}, whose statement holds
for any almost-sure bound on the corrected stream, the generic
$B_X+2KV$ of part~(ii) being one such choice.  Any bound violation aborts before CS
construction.  EB-CS excludes the exact mean in 2/400 raw replications and
0/400 AIVAT replications; AsympCS exclusions are 15/400 and 9/400.  At
$t=1000$ and 4000, the AIVAT EB-CS half-widths under $B_Y=117$ are 1.730 and
0.435, with empirical-to-floor ratios 1.0019 and 1.0074.  The historical
sample-derived $B=22$ execution remains a separate descriptive archive.

The AsympCS amendment uses 400 calibration and 400 evaluation replications
with disjoint seed ranges.  Calibration searches an 18-candidate grid over the
internal level, $t_{\rm opt}$, and the burn-in, and freezes a single candidate
before a one-time
evaluation.  Held-out exclusion counts for raw, frozen, predictable, and
oracle are 8, 5, 7, and 5 out of 400.  The corresponding four-arm
Bonferroni-adjusted one-sided Clopper--Pearson upper bounds are 4.21\%, 3.16\%,
3.86\%, and 3.16\%.  This is finite experimental calibration evidence and does
not convert AsympCS into a nonasymptotic method.

\subsection{A2 design details}

Each HUNL run is centered by its own AIVAT sample mean before the 15
sequences are concatenated into a pool of 71,439 values.  Each simulated entry
samples 5,000 values with replacement.  A fixed-sample normal interval is
monitored from $t=30$ and reports its first exclusion of zero.  The audit checks
EB-CS and AsympCS on the same prefix at that selected time.  The $H_0$ arm uses
2,000 entries; each injected-effect arm uses 500 and adds 0.05, 0.1, or 0.2 BB
per hand.

The A2 EB analysis declares a separate bound for every simulated entry as the
ceiling of that entry's observed maximum absolute payoff, with a floor of one.
It therefore uses future values beyond an early reported stop and is
sample-dependent.  Its 0/2,000 standalone detections and 1,227/1,227 selected-
stop containments are descriptive resampling outcomes, not formal
finite-sample coverage evidence.

\section{Robustness and Sensitivity Analyses}
\label{app:robustness}

These analyses were added after the main experiments and are descriptive.  They
address sensitivity to leaving out one HUNL run and to the length of a
simulated A2 leaderboard entry.  They do not repair the data-dependent-bound
limitations described in Appendix~\ref{app:experimental-details}.

\subsection{Leave-one-run-out HUNL summaries}

For each of the 15 HUNL runs in turn, we remove that complete run and recompute
the median over available run-level summaries among the remaining 14 runs.
Across the 15
leave-one-out datasets, the median variance-ratio summary ranges from 51.08 to
55.92, the $\pm1$ BB AsympCS stopping-ratio summary from 73.31 to 74.71, and the
corresponding EB-CS summary from 1.359 to 1.371.  Thus no single archived run
accounts for the large contrast between the two stopping methods.

\begin{table}[ht]
\centering
\small
\begin{tabular}{@{}lcc@{}}
\toprule
Quantity & Median over omissions & Range over omissions \\
\midrule
Variance ratio & $52.59\times$ & [$51.08$, $55.92$] \\
AsympCS stop ratio, $\pm1$ BB & $73.85\times$ & [$73.31$, $74.71$] \\
EB-CS stop ratio, $\pm1$ BB & $1.365\times$ & [$1.359$, $1.371$] \\
AIVAT AsympCS FP median, LOO splits & 9.6\% & [9.3\%, 10.3\%] \\
\bottomrule
\end{tabular}
\caption{Post hoc leave-one-run-out sensitivity.  Each entry first summarizes
available run-level values and then summarizes those 15 omission results.  The
EB stopping-ratio splits contain 13 or 14 evaluable ratios because one
full-data ratio is censored.}
\label{tab:loo-robustness}
\end{table}

The last row is the median across the 15 omission-specific medians of this
sensitivity analysis, not a mean false-positive rate.  It should not be confused
with P0's 7.1067\% aggregate: the latter averages both raw and AIVAT rates
across 30 conditions.

\subsection{A2 entry-length sensitivity}

\begin{figure}[t]
\centering
\includegraphics[width=0.72\linewidth]{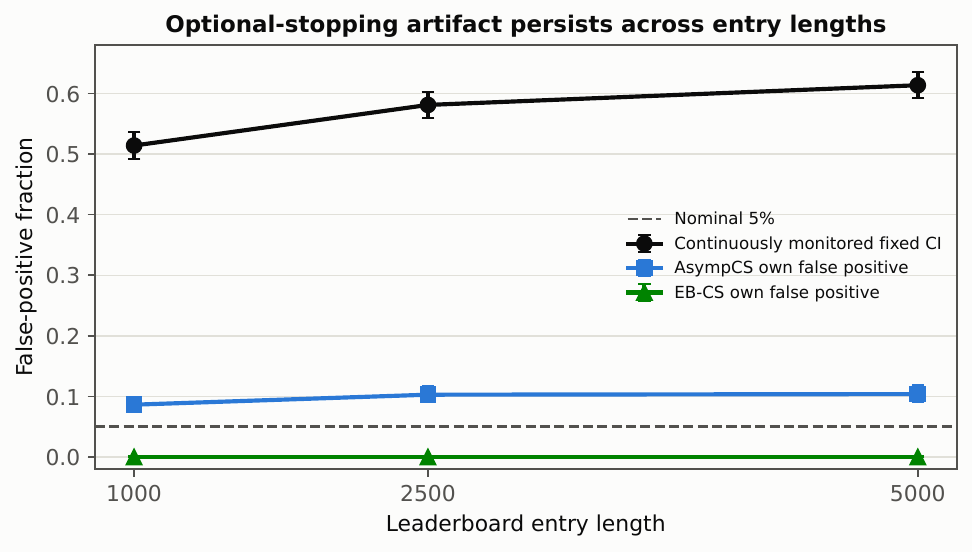}
\caption{Post hoc entry-length sensitivity.  Error bars are Wilson 95\%
intervals from the archived 2,000-entry results.}
\label{fig:a2-entry-length}
\end{figure}

We repeat the constructed $H_0$ audit with 2,000 entries at lengths 1,000,
2,500, and 5,000, using the same centered payoff pool and seed.  Longer entries
give the repeatedly monitored fixed interval more opportunities to cross: its
claim rate rises from 51.4\% to 61.35\%.  The audited AsympCS selected-stop
containment remains at least 99.90\%, while its standalone false-positive rate
is 8.65--10.4\%.

\begin{table}[ht]
\centering
\small
\begin{tabular}{@{}lrrrrr@{}}
\toprule
$N$ & Claims & Median $\hat t$ & EB contain & Asymp contain & Asymp FP \\
\midrule
1,000 & 1,028/2,000 & 52 & 100.0\% & 99.90\% & 8.65\% \\
2,500 & 1,162/2,000 & 64 & 100.0\% & 100.0\% & 10.3\% \\
5,000 & 1,227/2,000 & 59 & 100.0\% & 99.9185\% & 10.4\% \\
\bottomrule
\end{tabular}
\caption{Post hoc A2 $H_0$ sensitivity to simulated entry length.  ``Contain''
means that the interval still contains zero at the monitor's selected stop;
``Asymp FP'' is standalone detection by the horizon.}
\label{tab:a2-length-robustness}
\end{table}

The EB column is numerically stable but cannot be read as exact coverage:
every entry again uses the ceiling of its own observed maximum absolute payoff
as the bound.  The AsympCS column is an approximate screen, with finite-horizon
false-positive rates above the nominal 5\% in all three horizons.

\subsection{Controlled Leduc cross-settings}

A separate controlled extension uses one frozen policy pool and keeps the
reported payoff in seat 0.  S1 pairs the CFR$^+$-2000 policy with the uniform
opponent model used by the frozen evaluator.  Frozen and oracle corrected
streams are identical in this matched control, so there is no identifiable
oracle variance gap.  S2 pairs CFR$^+$-2000 with CFR$^+$-20; S3 reverses these
policy-budget roles.  The iteration labels identify construction budgets, not
a universal skill ordering.

In S2 and S3, predictable learning recovers 78.6\% and 77.1\% of the
frozen-to-oracle variance gap.  At the $\pm0.1$-chip target, its median stopping
times are 2240.5 and 2608 hands, compared with 3039 and 4337.5 for the frozen
arm and 1744 and 1425.5 for the oracle.  Raw streams are fully right-censored
at 8192 hands.  These are two specific opponent-model mismatch constructions,
not arbitrary-opponent or external-platform replication.

\subsection{Limits of the checks}

Centering each run constructs a zero-mean empirical pool but discards between-
run mean differences.  Sampling individual hands with replacement also removes
serial dependence and any run-level clustering within a simulated entry.
Consequently the A2 analysis demonstrates the effect of repeated peeking under
this resampling design; it does not estimate the error rate of every deployed
leaderboard process.  Likewise, leave-one-run-out stability addresses influence
of a single archived run, not external replication across opponents,
evaluators, or platforms.

\section{Assumption Checklist}
\label{app:arx-assumption-checklist}

Table~\ref{tab:arx-assumptions} collects the conditions AV-AIVAT relies on, the
statement each one supports, and the artifact that lets a reader confirm it.

\begin{table}[h]
\centering
\small
\begin{tabular}{@{}p{0.24\textwidth}p{0.26\textwidth}p{0.40\textwidth}@{}}
\toprule
Assumption & Needed for & Check or report \\
\midrule
Independent bound $B_Y$ &
Exact EB-CS certificate &
The HUNL observed-containment check is descriptive; the controlled Leduc run
uses the analytic $B_Y=117$ certificate of
Appendix~\ref{app:experimental-details}. \\
Predictable $v_t$ &
Mean-zero corrections &
Algorithm~\ref{alg:arx-protocol} updates only $v_{t+1}$ after hand $t$;
Section~\ref{sec:exp-e2} uses past hands only. \\
Known conditional kernel &
AIVAT unbiasedness &
Log $p_{t,h}(\cdot)$ given the complete pre-action information at every enabled
chance or evaluated-agent correction node. \\
Pre-action enablement &
AIVAT unbiasedness &
Record reached and enabled flags; $S_{t,h}$ must be fixed before the outgoing
action and the later trajectory. \\
No current-hand peeking &
Legal online learning &
Algorithm~\ref{alg:arx-protocol} fixes $v_t$ before hand $t$; only $v_{t+1}$
may use it (Section~\ref{sec:arx-validity}). \\
AsympCS approximation &
Practical interval &
Finite-sample miscoverage is reported in every experiment of
Section~\ref{sec:experiments}. \\
Paired log release &
Reproducibility &
Rechecking a claim uses the $Y$ prefix and stop metadata; paired $(X_t,Y_t)$ is
an optional raw-stream release (Section~\ref{sec:release-protocol}). \\
\bottomrule
\end{tabular}
\caption{Checklist connecting the AV-AIVAT assumptions to the reported checks.}
\label{tab:arx-assumptions}
\end{table}

\section{Research Scope}
\label{app:scope}

AV-AIVAT targets sequential evaluations with a verifiable conditional
action kernel or another conditional mean-zero correction mechanism.  The HUNL
evidence covers 15 configurations, not necessarily distinct base models, on one
paired-metric platform.  E2 supplies one matched-prior control and two mismatched
Leduc policy pairs; its 77--79\% variance-gap reduction characterizes those
controlled constructions.  A2 resamples individual centered hands and therefore
studies optional stopping without preserving temporal dependence or run-level
clustering.  The same interface can be applied whenever the correction law is
verifiable; extending the empirical evidence to additional platforms,
dependent streams, multiplayer calibration, and per-seat analysis requires the
corresponding data and sampling model.  Reconstructing a stopping claim from its
released prefix and metadata is distinct from detecting a manipulated correction
generator.

\section{Future Work}
\label{app:future-work}
Several extensions would change the evidence base rather than enlarge the
current experiment.  First, finite-sample intervals designed for heavy-tailed or
adaptively truncated payoffs should be compared on the same streams using
coverage, width, and stopping cost jointly.  A narrower interval without a valid
bias or tail argument is not a substitute for the EB theorem.

Second, concrete variance-regret rates are needed for tabular opponent models and
neural value learners.  In games, the least-visited relevant information set can
control sup-norm value error even when average prediction error improves rapidly.
A visitation-aware analysis could relate reach probabilities, model estimation,
and finite-horizon stopping directly.

Third, a larger audit benchmark should inject isolated state, sign, seat, cache,
node-selection, and update-order faults and preregister diagnostic thresholds.
The two current defect logs show that simple checks can reveal severe failures;
they do not rank diagnostics under a representative fault distribution.

Finally, multi-platform and multiplayer studies are needed.  Cross-platform data
would separate evaluator-specific behavior from general statistical phenomena.
Multiplayer evaluation requires explicit seat conditioning, node-law accounting,
and dependence analysis before the two-player protocol can be transferred.

\end{document}